\documentclass[12pt]{article}
\usepackage{epsf}
\usepackage{amsmath}
\def\hybrid{\topmargin 0pt      \oddsidemargin 0pt
	\headheight 0pt \headsep 0pt
	\textheight 9in         
	\textwidth 6.25in       
	\marginparwidth .875in
	\parskip 5pt plus 1pt   \jot = 1.5ex}

\catcode`\@=11
\def\marginnote#1{}
\newcount\hour
\newcount\minute
\newtoks\amorpm
\hour=\time\divide\hour by60
\minute=\time{\multiply\hour by60 \global\advance\minute by-\hour}
\edef\standardtime{{\ifnum\hour<12 \global\amorpm={am}%
	\else\global\amorpm={pm}\advance\hour by-12 \fi
	\ifnum\hour=0 \hour=12 \fi
	\number\hour:\ifnum\minute<10 0\fi\number\minute\the\amorpm}}
\edef\militarytime{\number\hour:\ifnum\minute<10 0\fi\number\minute}

\def\draftlabel#1{{\@bsphack\if@filesw {\let\thepage\relax
   \xdef\@gtempa{\write\@auxout{\string
      \newlabel{#1}{{\@currentlabel}{\thepage}}}}}\@gtempa
   \if@nobreak \ifvmode\nobreak\fi\fi\fi\@esphack}
	\gdef\@eqnlabel{#1}}
\def\@eqnlabel{}
\def\@vacuum{}
\def\draftmarginnote#1{\marginpar{\raggedright\scriptsize\tt#1}}

\def\draft{\oddsidemargin -.5truein
	\def\@oddfoot{\sl preliminary draft \hfil
	\rm\thepage\hfil\sl\today\quad\militarytime}
	\let\@evenfoot\@oddfoot \overfullrule 3pt
	\let\label=\draftlabel
\let\marginnote=\draftmarginnote
	\let\marginnote=\draftmarginnote
   \def\@eqnnum{(\theequation)\rlap{\kern\marginparsep\tt\@eqnlabel}%
\global\let\@eqnlabel\@vacuum}  }

\def\numberbysection{\@addtoreset{equation}{section}
\def\theequation{\thesection.\arabic{equation}}}

\def\underline#1{\relax\ifmmode\@@underline#1\else
	$\@@underline{\hbox{#1}}$\relax\fi}

\def\titlepage{\@restonecolfalse\if@twocolumn\@restonecoltrue\onecolumn
     \else \newpage \fi \thispagestyle{empty}\c@page\z@
	\def\thefootnote{\fnsymbol{footnote}} }

\def\endtitlepage{\if@restonecol\twocolumn \else  \fi
	\def\thefootnote{\arabic{footnote}}
	\setcounter{footnote}{0}}  
\catcode`@=12
\relax

\def\beq{\begin{equation}}
\def\eeq{\end{equation}}
\def\bea{\begin{eqnarray}}
\def\eea{\end{eqnarray}}
\def\nn{\nonumber}

\relax
\hybrid

\begin{document}

\begin{titlepage}
\begin{center}
June~2016 \hfill . \\[.5in]
{\large\bf Correlation function of four spins in the percolation model. }
\\[.5in] 
{\bf Vladimir S.~Dotsenko}\\[.2in]
{\it LPTHE, CNRS, Universit{\'e} Pierre et Marie Curie, Paris VI, UMR 7589\\
               4 place Jussieu,75252 Paris Cedex 05, France.}\\[.2in]
               
 \end{center}
 
\underline{Abstract.}

By using the Coulomb gas technics we calculate 
the four-spin correlation function in the percolation $q\rightarrow 1$ 
limit of the Potts model.

It is known that the four-point functions define the actual fusion rules of a particular model. In this respect, we find that fusion of two spins, of dimension $\Delta_{\sigma}=\frac{5}{96}$, produce a new channel, in the 4-point function, which is due to the operator with dimension $\Delta=5/8$.

\end{titlepage}

\newpage

\numberwithin{equation}{section}

\section{Introduction.}

In the renewed interest to the Potts model correlation functions, 
spin and cluster connectivity functions [1,2], there has been considerable progress recently
in defining the spin and connectivity three-point functions, for the Potts model with $q$ general, 
$q<4$, [3,1,4], and also for the corresponding three-point functions
of the loop models [5].

En the other hand, defining multipoint functions, starting with the four-point spin and connectivity functions for the Potts model with $q$ general, this problem presents, for the moment, considerable difficulties.

In the present paper we shall present a limited progress in that direction, 
by defining, analytically, the conformal theory four-spin correlation function 
for the percolation limit of the Potts model, $q\rightarrow 1$.
This function exhibits a new channel, due to the operator with dimension 
$\Delta=5/8$, produced by fusion of two spins.

With respect to the 4-spin function defined on the lattice, we suppose that our conformal theory function is a particular bloc in a finite linear combination of other 4-point functions. This is because the lattice spin operator, in general, is a linear combination of a leading conformal theory spin operator plus the subleading ones. This is except for very simple models, like Ising model.

As a consequence, the lattice spin 4-point function will break into a linear combination of different conformal theory proper operators functions (proper with respect to $L_{0}$), of the leading and subleading spins. As fusion of subleading spins (or, of a leading and a subleading spins) may produce, 
back, the leading spin operator, along with the other
intermediate channels, the 4-point functions will get mixed.

This is different from the case of 3-point functions, in which a product of two lattice spin operators is projected onto another spin operator, formally placed at infinity. In the corresponding limiting procedure the 3-point function of leading conformal theory spins will get neatly separated, with the appropriate coefficient, resulting into one to one correspondence between the conformal theory and lattice model 3-point functions.

This is different for the 4-point functions, for the reasons raised above. As a result the correspondence between the 4-point functions defined in the conformal theory, for proper operators, and the corresponding functions defined on lattice, this correspondence gets complicated.

In relatively simple cases it is possible to define proper operators 
directly out of  the lattice ones, by doing a particular Fourier analysis, for instance. But in general, defining the subleading proper operators (to get them separated) directly on the lattice, this might be very complicated.

With respect to the function that we have calculated, for proper spin operators, it should not be expected, on general grounds, that it could
 be expressed as a simple linear combination of 4-point lattice spin functions, 
or 4-point cluster connectivities, defined in [2].

But, on the other hand, the new channel that we have found, of dimension 
$\Delta=5/8$, it should be present in the 4-point symmetric cluster connectivity functions of [2], as a sub-leading channel. The leading channel 
appears to be taken by the spin operator itself, being produced, as we suppose, by particular cross-fusions with subleading spin operators.  

Some further remarks of a similar nature, but presented
somewhat differently, are given at the end of the Section 3, with one additional simple example, the exemple which could be treated with the Fourier analysis.

Saying it again, in the present paper we suppose that we have defined one particular bloc of the lattice 4-spin function, the bloc which is accessible 
by the Coulomb gas technics. Defining the other blocs and the full lattice 4-spin function, together with the corresponding 4-point connectivities, 
remains an open problem. On the conformal theory side this amounts 
to defining the full operator algebra generated by the conformal 
spin operators.

\section{General presentation of the method.}

In the minimal model context of the conformal field theory [6], the spin 
operator of the Potts model is represented by the primary fields [7]:
\beq
\Phi_{\frac{p+1}{2},\frac{p-1}{2}}, \quad \mbox{or, equivalently,} \quad  \Phi_{\frac{p+1}{2},\frac{p+1}{2}} \label{eq2.1}
\eeq
Here $p$ is the parameter of the minimal model $M_{p}$ having the central charge
\beq
c\equiv c_{p}=1-\frac{6}{p(p+1)} \label{eq2.2}
\eeq

In the Coulomb Gas representation [8], the primary fields (\ref{eq2.1}) are represented by the vertex operators
\bea
V_{\alpha_{\sigma}}(z,\bar{z})=e^{i\alpha_{\sigma}\varphi(z,\bar{z})}\nn\\
\mbox{with} \quad \alpha_{\sigma}=\alpha_{\frac{p+1}{2},\frac{p-1}{2}}, 
\quad \mbox{or, equivalently}, \quad \alpha_{\sigma}=\alpha_{\frac{p+1}{2},\frac{p+1}{2}}
\label{eq2.3}
\eea
($\sigma$ stands for the spin operator). Here the Coulomb Gas charges $\{\alpha_{n',n}\}$ are defined as:
\bea
\alpha_{n',n}=\frac{1-n'}{2}\alpha_{-}+\frac{1-n}{2}\alpha_{+}\nn\\
\alpha_{+}=\sqrt{\frac{p+1}{p}},\quad\alpha_{-}=-\sqrt{\frac{p}{p+1}} 
\label{eq2.4}
\eea

The multipoint correlation functions of primary operators (or fields) are given by the multiple integrals (presented symbolically):
\bea
<V_{1}V_{2}V_{3}V_{4}>_{conf.}\nn\\
\propto<V_{1}V_{2}V_{3}V_{4}(\int d^{2}uV_{-}(u,\bar{u}))^{l}(\int d^{2}vV_{+}(v,\bar{v}))^{k}> \label{eq2.5}
\eea
In the l.h.s. we have the conformal field theory correlation function of four fields (just as an example) $V_{1},V_{2},V_{3},V_{4}$.
In the r.h.s. we have the average, over the free field $\varphi(z,\bar{z})$, 
of vertex operators $V_{1},V_{2},V_{3},V_{4},V_{-},V_{+}$, defined 
as in (\ref{eq2.3}). $V_{-},V_{+}$ are the screening operators:
\bea
V_{-}(u,\bar{u})=e^{i\alpha_{-}\varphi(u,\bar{u})},\nn\\
V_{+}(v,\bar{v})=e^{i\alpha_{+}\varphi(v,\bar{v})} \label{eq2.6}
\eea
(with $\alpha_{-}$, $\alpha_{+}$ in (\ref{eq2.4})) having conformal dimensions 
$\Delta_{-}=\Delta_{+}=1$.

The numbers $l,k$, of the integrals in (\ref{eq2.5}), which are numbers of screening 
operators (\ref{eq2.6}) integrated over the whole 2D plane, 
are defined by the neutrality condition
\beq
\alpha_{1}+\alpha_{2}+\alpha_{3}+\alpha_{4}+l\alpha_{-}+k\alpha_{+}=2\alpha_{0} \label{eq2.7}
\eeq
Here $\alpha_{1}$, $\alpha_{2}$, $\alpha_{3}$, 
$\alpha_{4}$ are the Coulomb Gas charges of the operators 
$V_{1}$, $V_{2}$, $V_{3}$, $V_{4}$; $\alpha_{0}$ is the background charge of the Coulomb Gas. In particular
\beq
2\alpha_{0}=\alpha_{-}+\alpha_{+} \label{eq2.8}
\eeq

More details, with respect to the formulas 
(\ref{eq2.3}) - (\ref{eq2.8}) could be found in [8]. We have reproduced these standard formulas just for close references, because we are going to use this representation in our calculations.

Evidently, the integral representation (\ref{eq2.5}), 
for the correlation functions of four operators, 
could be used if the numbers of screenings, $l$ and $k$, 
defined by the neutrality condition (\ref{eq2.7}), are positive integers. 
Otherwise the correlation function has to be defined by the analytic continuation. The analytic continuation is known for 3-point functions, [3,4,9] and references there-in. It is not known at present for the 4-point functions.

For the 4-spin correlation function of the Potts model, with the corresponding vertex operator in (\ref{eq2.3}), 
\beq
<\sigma\sigma\sigma\sigma>=
<V_{\alpha_{\sigma}}V_{\alpha_{\sigma}}V_{\alpha_{\sigma}}V_{\alpha_{\sigma}}>_{conf.} 
\label{eq2.9}
\eeq
will shall have, by the condition (\ref{eq2.7}), the numbers $l,k$ being, in general, non-integer, except for special values of the parameter $p$: $p=3$, Ising model, $p=5$, $q=3$ Potts model, $p=2m+1$, $m=3,4,5,...$, for higher minimal models.

For the above sequence of values of $p$, the two indices of the spin operator, $\frac{p+1}{2}$ \\ and $\frac{p-1}{2}$ , eq.(\ref{eq2.3}), will have integer values, corresponding to primary operators with degenerate representations, of the corresponding minimal models.

For the percolation limit, $q\rightarrow 1$, the corresponding conformal field theory has
\beq
p=2 \label{eq2.10}
\eeq
- the central charge (\ref{eq2.2}) is zero, and the spin operator is represented
by the fields (\ref{eq2.1}):
\beq
\Phi_{\frac{3}{2},\frac{1}{2}}(z,\bar{z}),\quad \mbox{or} \quad
\Phi_{\frac{3}{2},\frac{3}{2}}(z,\bar{z}) \label{eq2.11}
\eeq
or by the vertex operators $V_{\alpha_{\sigma}}(z,\bar{z})$, with
\beq
\alpha_{\sigma}=\alpha_{\frac{3}{2},\frac{1}{2}}=-\frac{\alpha_{-}}{4}+\frac{\alpha_{+}}{4} \label{eq2.12}
\eeq
or
\beq
\alpha_{\sigma}=\alpha_{\frac{3}{2},\frac{3}{2}}=-\frac{\alpha_{-}}{4}-\frac{\alpha_{+}}{4} \label{eq2.13}
\eeq

 The indices of the fields (\ref{eq2.11}) are non-integer. 
 They are not the usual primary fields of the minimal models, 
 they are non-degenerate. Their correlation functions have to be defined, in principal, by the analytic continuation, in their indices, from the integer values.
 
But if we define the four-point function by the Coulomb Gas representation
\beq
<\sigma\sigma\sigma\sigma>\sim<V_{\alpha_{\sigma}}V_{\alpha_{\sigma}}V_{\alpha_{\sigma}}V_{\alpha_{\sigma}}>_{conf.} \label{eq2.14}
\eeq
with $\alpha_{\sigma}$ in (\ref{eq2.12}) or in (\ref{eq2.13}), 
we find that, by (\ref{eq2.7}), the numbers of screenings $l$ and $k$ are integers, so that  the correlation function of four spins could be defined by the integral representation in (\ref{eq2.5}).

In particular, if $\alpha_{\sigma}$ is taken as in (\ref{eq2.12}), 
for the function in (\ref{eq2.14}), then one finds, by (\ref{eq2.7}), that
\beq
l=2,\quad k=0 \label{eq2.15}
\eeq

If $\alpha_{\sigma}$ is taken as in (\ref{eq2.13}), which should be equivalent, we find
\beq
l=2,\quad k=2 \label{eq2.16}
\eeq

In the next Section, we shall justify that the numbers of intermediate 
channels (the numbers of conformal blocks), for the four-point function 
defined by the integral representation (\ref{eq2.5}), is 2, 
for $l$ and $k$ in (\ref{eq2.15}), and is 5, for $l$ and $k$ 
in (\ref{eq2.16}). But, somehow, if we are still dealing with the conformal theory of the Potts model, the two different representations should give the same four-point function. This consistency will be checked. For the moment we are still in the process of defining our problem, or problems.

\underline{One additional comment.}

Usually, for the integral representation of 4-point correlation functions, in the case of minimal models, the preferred representation, which is equivalent but which requires minimal numbers of screenings (of integrations) 
is that of [8]
\beq
<V^{+}_{\alpha_{\sigma}}V_{\alpha_{\sigma}}V_{\alpha_{\sigma}}V_{\alpha_{\sigma}}>_{conf.}=<V_{\alpha^{+}_{\sigma}}V_{\alpha_{\sigma}}V_{\alpha_{\sigma}}V_{\alpha_{\sigma}}>_{conf.} \label{eq2.17}
\eeq
-- with one of the vertex operators taken in the conjugate representation, with the charge
\beq
\alpha^{+}_{\sigma}=2\alpha_{0}-\alpha_{\sigma} \label{eq2.18}
\eeq
But in the present case, with the spin operator having 
non-integer indices, combining the vertex operators as in (\ref{eq2.17}) would lead to the integral representation 
with non-integer values of $l$ and $k$, 
by the condition (\ref{eq2.7}), so that it cannot be used. The function should be defined by the analytic continuation, which should provide the same function, but for the moment the analytic continuation 
is not known, for 4-point functions.

The way out, which exist at the percolation point, 
only at this point and not in its vicinity, is to use the representation in (\ref{eq2.9}), without the conjugate vertex operators, with $\alpha_{\sigma}$ either in (\ref{eq2.12}) 
or in (\ref{eq2.13}).

One particular conclusion, which is due to the existence of the integral representation, will be that the spectrum of intermediate channels, of the four-point function $<\sigma\sigma\sigma\sigma>$, is discrete, for the percolation problem.

We shall calculate the corresponding function, by (\ref{eq2.5}), 
in the next Section.

\section{Calculation of the function $<\sigma\sigma\sigma\sigma>$.}

To calculate the function
\beq
<\sigma(\infty)\sigma(1)\sigma(z,\bar{z})\sigma(0)> \label{eq3.1}
\eeq
we shall use the Coulomb Gas representation (\ref{eq2.5}) with
\beq
V_{1}=V_{2}=V_{3}=V_{4}=V_{\alpha_{\sigma}} \label{eq3.2}
\eeq
and
\beq
\alpha_{\sigma}=\alpha_{\frac{3}{2},\frac{1}{2}}=-\frac{\alpha_{-}}{4}+\frac{\alpha_{+}}{4} \label{eq3.3}
\eeq
This representation requires a minimal number of integrations, $l=2$, $k=0$, eq.(\ref{eq2.15}).

The verification that the representation 
with $\alpha_{\sigma}=\alpha_{\frac{3}{2},\frac{3}{2}}$ (and numbers of screenings $l=2$, $k=2$, eq.(\ref{eq2.16})) gives the same function, this verification will be done in the next Section. 

One finds:
\bea
<\sigma(\infty)\sigma(1)\sigma(z,\bar{z})\sigma(0)>\equiv G(z,\bar{z})\nn\\
\propto<V_{\alpha_{\sigma}}(\infty)V_{\alpha_{\sigma}}(1)V_{\alpha_{\sigma}}(z,\bar{z})V_{\alpha_{\sigma}}(0)>_{conf.}\nn\\
\propto\int d^{2}u_{1}\int d^{2}u_{2}
<V_{\alpha_{\sigma}}(\infty)V_{\alpha_{\sigma}}(1)V_{\alpha_{\sigma}}(z,\bar{z})V_{\alpha_{\sigma}}(0)V_{-}(u_{1},\bar{u}_{1})V_{-}(u_{2},\bar{u}_{2})>
\label{eq3.4}
\eea

Accordingly to the techniques of [8,10], presented with some more details 
in [11], the function (the integral) in the r.h.s. of (\ref{eq3.4}) 
could be factorized as:
\beq
G(z,\bar{z})\propto\sum_{i=0,1,2}\frac{1}{|z|^{4\Delta_{\alpha}-2\Delta_{p_{i}}}}
\times C_{\alpha,\alpha,p_{i}}C_{p^{+}_{i},\alpha,\alpha}|F_{p_{i}}(z)|^{2}
\label{eq3.5}
\eeq
Here $\alpha\equiv\alpha_{\sigma}=-\frac{\alpha_{-}}{4}+\frac{\alpha_{+}}{4}$;
\beq
\{p_{i},i=0,1,2,\}=\{p_{0}=2\alpha, \,\,\, p_{1}=2\alpha+\alpha_{-}, \,\,\, p_{2}=2\alpha+2\alpha_{-}\} \label{eq3.6}
\eeq
are the charges of the intermediate channels ($p_{i}$, here, 
have nothing to do with the parameter $p$ of minimal models, 
eq.(\ref{eq2.1}), (\ref{eq2.2}));
\beq
C_{p^{+}_{i},\alpha,\alpha}=<V_{p^{+}_{i}}(\infty)V_{\alpha}(1)V_{\alpha}(0)>_{conf.} \label{eq3.7}
\eeq
\beq
C_{\alpha,\alpha,p_{i}}=<V_{\alpha}(\infty)V_{\alpha}(1)V_{p_{i}}(0)>_{conf.}
\label{eq3.8}
\eeq
are the Coulomb Gas structure constants. The 3 point function $C_{p^{+}_{i},\alpha,\alpha}$ in (\ref{eq3.7}) is actually the product of the first two operators $V_{\alpha}(z,\bar{z})V_{\alpha}(0)$, in the four-point function (\ref{eq3.4}), projected ento $V^{+}_{p_{i}}(\infty)$, in order to pick up the first term of the expansion
\beq
V_{\alpha}(z,\bar{z})V_{\alpha}(0)=\frac{C^{p_{i}}_{\alpha,\alpha}}{|z|^{4\Delta_{\alpha}-2\Delta_{p_{i}}}}V_{p_{i}}(0)+...
\label{eq3.8A}
\eeq
This gives the correlation function 
$<V_{p_{i}^{+}}(\infty)V_{\alpha}(z,\bar{z}) V_{\alpha}(0)>_{conf.}$, 
and then the $|z|$ factor is ruled out, as in (\ref{eq3.5}), (\ref{eq3.7}). 
We observe also that
\beq
C^{p_{i}}_{\alpha,\alpha}=C_{p^{+}_{i},\alpha,\alpha} \label{eq3.8B}
\eeq

$\{F_{p_{i}}(z)\}$ are the conformal block functions of the corresponding channels. They are supposed to be normalised by 1:
\beq
\mbox{as} \,\, z\rightarrow 0, \quad F_{p_{i}}(z)\rightarrow 1 \label{eq3.8C}
\eeq
i.e.
\beq
F_{p_{i}}(z)=1+k_{1}z+k_{2}z^{2}+...
\label{eq3.8D}
\eeq

\underline{We remind the technics of [8,10].}

It is known that the 2D integral in the r.h.s. of (\ref{eq3.4}) 
could be expressed as a sum of 1D modulus squared holomorphic (in $z$) integrals. 
In this sum, of modulus squared contour integrals, the different terms are classified by the distribution of the contours of integration: in the first term, no contours between $0$ and $z$, they are all put 
between $1$ and $\infty$; in the second term, one contour (one screening) is integrated between $0$ and $z$, the rest, between $1$ and $\infty$, and so on. In this way one gets the sum in (\ref{eq3.5}), $F_{p_{i}}(z)$ being the corresponding contour integrals. 

The coefficients of this sum factorize ento 
the constants $C_{\alpha,\alpha,p_{i}}$ and $C_{p^{+}_{i},\alpha,\alpha}$ which could, equivalently, be defined directly, by the corresponding distributions of the $2D$ integrations: $0$ and $z, \bar{z}$ could completely be separated from $1$ and $\infty$, when $z,\bar{z}\rightarrow 0$, 
and the screenings, their  $2D$ integrations, could be distributed accordingly. For some more details see also [11].

The 3-point functions in (\ref{eq3.7}), (\ref{eq3.8}) 
correspondent to the limiting factorization of the 4-point function, as $z,\bar{z}\rightarrow 0$:

1) for the channel $p_{0}=2\alpha$, no screenings are present around $(0,(z,\bar{z}))$, i.e. no screenings in the 3-point function $C_{p^{+}_{0}\alpha,\alpha}$; both screenings are being put 
into the 3-point function $C_{\alpha,\alpha,p_{0}}$;

2) for the channel $p_{1}=2\alpha + \alpha_{-}$, one screening is in $C_{p^{+}_{1},\alpha,\alpha}$ (around $0,(z,\bar{z})$) 
and an another one is in $C_{\alpha,\alpha,p_{1}}$;

3) for the channel $p_{2}=2\alpha+2\alpha_{-}$, both screenings are in $C_{p^{+}_{2},\alpha,\alpha}$, being integrated around $(0,(z,\bar{z}))$, and no screenings in $C_{\alpha,\alpha,p_{2}}$.

In this way, by factorizing the Coulomb Gas integral in (\ref{eq3.4}), one defines the intermediate channels of the 4-point function.

The corresponding integrals, for $C_{p^{+},\alpha,\alpha}$ and $C_{\alpha,\alpha,p}$ have been evaluated in [10]. Somewhat more symmetric form, for the structure constants 
$C_{\alpha_{1},\alpha_{2},\alpha_{3}}$, is given in [9], eq.(4.8), the expression which we shall use here, with just a factor $\frac{1}{Z}$ to be added to the expression in (4.8), according to our analysis of normalisations in the Section 4 of [9]. $Z$ is the Coulomb Gas partition function [9].

It is easily seen that, for the symmetric 4-point function in (\ref{eq3.4}), (\ref{eq3.5}), the channels $p_{0}$ and $p_{2}$ are identical. In particulur, $p^{+}_{2}=2\alpha_{0}-p_{2}=p_{0}$, 
$p^{+}_{0}=2\alpha_{0}-p_{0}=p_{2}$, so that the channel $p_{0}$ would just appear twice in the decomposition (\ref{eq3.5})

The  independent channels in (\ref{eq3.5}) are $p_{0}$ and $p_{1}$, with:
\bea
p_{0}=2\alpha=-\frac{\alpha_{-}}{2}+\frac{\alpha_{+}}{2},\nn\\
\Delta_{p_{0}}=(p_{0}-\alpha_{0})^{2}-\alpha^{2}_{0}=\frac{5}{8}\label{eq3.9}
\eea
\bea
p_{1}=2\alpha+\alpha_{-}=\frac{\alpha_{-}}{2}+\frac{\alpha_{+}}{2}
=\alpha_{0} \nn\\
\Delta_{p_{1}}=(p_{1}-\alpha_{0})^{2}-\alpha^{2}_{0}
=-\frac{1}{24}\label{eq3.10}
\eea

We remind that, for the percolation,
\beq
\Delta_{\sigma}=\Delta_{\alpha}=(\alpha-\alpha_{0})^{2}-\alpha^{2}_{0}=\frac{5}{96} \label{eq3.11}
\eeq
$\alpha\equiv\alpha_{\sigma}$ is given in (\ref{eq3.3}).
We remind also that
\bea
\alpha_{+}=\sqrt{\frac{3}{2}},\quad\alpha_{-}=-\sqrt{\frac{2}{3}}\nn\\
\alpha_{0}=\frac{\alpha_{+}+\alpha_{-}}{2}=\frac{1}{\sqrt{24}}=\frac{1}{2\sqrt{6}}\label{eq3.12}
\eea

We shall see shortly that the channel $p_{1}$, with the negative dimension 
of the intermediate operator, actually decouples, due to particular values of the constants $C_{\alpha,\alpha,p_{0}}$, $C_{p^{+}_{0},\alpha,\alpha}$, $C_{\alpha,\alpha,p_{1}}$, $C_{p^{+}_{1},\alpha,\alpha}$ in (\ref{eq3.5}). There remains a single intermediate channel, in $<\sigma\sigma\sigma\sigma>$, 
for the percolation, the one with the dimension 
of the intermediate operator $\Delta_{p_{0}}=5/8$, eq.(\ref{eq3.9}).

The structure constants $C_{\alpha,\alpha,p_{i}}$, $C_{p^{+}_{i},\alpha,\alpha}$ are those for the operator algebra of the vertex operators, with their non-trivial normalisations [11,9]. Alternatively, the 4-point function (\ref{eq3.4}) could be decomposed as:
\beq
G(z,\bar{z})\propto\sum_{i=0,1,2}\frac{1}{|z|^{4\Delta_{\alpha}-2\Delta_{p_{i}}}}
\times(D_{\alpha, \alpha, p_{i}})^{2}\cdot|F_{p_{i}}(z)|^{2}\label{eq3.13}
\eeq
where the structure constants $D_{\alpha,\alpha,p_{i}}$ are those for the operators normalised by 1:
\beq
\Phi_{\alpha}=\frac{1}{N_{\alpha}}V_{\alpha},\quad \Phi_{P_{i}}=\frac{1}{N_{p_{i}}}V_{p_{i}}
\label{eq3.14}
\eeq
$N_{\alpha}$, $N_{p_{i}}$ are the norms of the vertex operators $V_{\alpha}$, $V_{p_{i}}$ [11.9]. 
The constants for the normalised operators, $D_{\alpha,\alpha,p_{i}}$, could be presented as in [3], 
eq.(5.1), in terms of $\Upsilon$ functions.
See also [9], eq.(4.41), where these constants are given in slightly different notations for the $\Upsilon$ functions, in the form which will be used here.

The decompositions (\ref{eq3.5}) and (\ref{eq3.13}) are equivalent.

The coefficients in these decompositions are related as:
\beq
C_{\alpha,\alpha,p_{i}}C_{p^{+}_{i},\alpha,\alpha}=\frac{(N_{\alpha})^{4}}{Z}(D_{\alpha,\alpha,p_{i}})^{2}\label{eq3.15}
\eeq
This is not difficult to justify, given the expressions of the coefficients $C_{\alpha,\alpha,p_{i}}$ and $C_{p^{+}_{i},\alpha,\alpha}$ in terms of the 3-point functions in (\ref{eq3.7}), (\ref{eq3.8}).

The only subtle point, one has to take into account that the normalisations of the operators $V_{p_{i}}$ and $V_{p^{+}_{i}}$ are given by ([9], eq.(4.34)):
\beq
N(V_{p_{i}})\equiv N_{p_{i}} \quad \mbox{and} \quad N(V_{p^{+}_{i}})\equiv N_{p_{i}^{+}}
=\frac{1}{Z \cdot N_{p_{i}}} 
\label{eq3.16}
\eeq
The expression for $N_{\alpha}$, $N_{p}$ is given in [9], eq.(4.37).

It could be remarked that the constants of the normalised operators $D_{\alpha,\alpha,p}$ are totally symmetric; they are symmetric also with respect to the conjugation:
\beq
D_{p^{+},\alpha,\alpha}=D_{p,\alpha,\alpha}=D_{\alpha,\alpha,p}\label{eq3.17}
\eeq

We remind again that to the expressions for the Coulomb Gas constants $C_{\alpha,\alpha,p}$, $C_{p^{+},\alpha,\alpha}$ in (\ref{eq3.15}), to the expression given in [9], eq.(4.8), has to be added 
the factor $1/Z$, $Z$ being the Coulomb Gas partiition function, defined in [9], eq.(4.11).

The equality (\ref{eq3.15}) has been tested and used many times in the course of calculations. In particular cases the calculation of constants $D$, expressed in $\Upsilon$ functions, is much easier than that using the expression (4.8), [9], for the Coulomb Gas the constants. 
In the other cases it is the opposite.

The conformal block functions $\{F_{p_{i}}(z)\}$ could be calculated by the standard algebra of the Virasoro descendants [6]. We shall give their series decomposition, 
in powers of $z$, up to the 4th order, $z^{4}$, just below.

Alternatively, these functions could be given by the contour integrals, exactly, valid for all values of $z$. The contour $(1D)$ integrals are obtained, as in [8], [10], from the $2D$ integrals in (\ref{eq3.4}). This integral representation, of the functions $F_{p_{i}}(z)$, will be given in the Appendix A.

Now we can give the explicit results of our calculations, of the function \\ $G(z,\bar{z})
=<\sigma(\infty)\sigma(1)\sigma(z,\bar{z})\sigma(0)>$, expressed as in (\ref{eq3.5}).

The expansion (\ref{eq3.5}), for the function
\bea
G(z,\bar{z})=<\sigma(\infty)\sigma(1)\sigma(z\bar{z})\sigma(0)>\nn\\
\propto<V_{\alpha}(\infty)V_{\alpha}(1)V_{\alpha}(z,\bar{z})V_{\alpha}(0)>_{conf.}\label{eq3.18}
\eea
with $\alpha=\alpha_{\sigma}$ in (\ref{eq3.3}), could be obtained by first developing, by the operator algebra, the product of the first two operators in (\ref{eq3.18}):
\bea
V_{\alpha}(z,\bar{z})V_{\alpha}(0)=\sum_{p}\frac{C^{p}_{\alpha,\alpha}}{|z|^{4\Delta_{\alpha}-2\Delta_{p_{i}}}}\nn\\
\times\{V_{p}(0)+z\beta^{(-1)}_{p}L_{-1}V_{p}(0)+z^{2}[\beta^{(-1,-1)}_{p}L^{2}_{-1}V_{p}(0)+\beta^{(-2)}_{p}L_{-2}V_{p}(0)]\nn\\
+z^{3}[\beta_{p}^{(-1,-1,-1)}L^{3}_{-1}V_{p}(0)+\beta_{p}^{(-1,-2)}L_{-1}L_{-2}V_{p}(0)+\beta_{p}^{(-3)}L_{-3}V_{p}(0)]\nn\\
+z^{4}[\beta_{p}^{(-1,-1,-1,-1)}L_{-1}^{4}V_{p}(0)+\beta_{p}^{(-1,-1,-2)}L_{-1}^{2}L_{-2}V_{p}(0)\nn\\
+\beta_{p}^{(-1,-3)}L_{-1}L_{-3}V_{p}(0)+\beta_{p}^{(-2,-2)}L_{-2}^{2}V_{p}(0)+\beta^{(-4)}_{p}L_{-4}V_{p}(0)]+...\}\label{eq3.19}
\eea 
and then substituting back this development into the function (\ref{eq3.18}) (we shall drop the precision "conf." for the type of the correlation function, in what follows; but it will always be assumed; in fact, the distinction was necessary only at the start, in eq.(\ref{eq2.5})):
\bea
<V_{\alpha}(\infty)V_{\alpha}(1)V_{\alpha}(z,\bar{z})V_{\alpha}(0)>=\sum_{p}\frac{C^{p}_{\alpha,\alpha}}{|z|^{4\Delta_{\alpha}-2\Delta_{p}}}\nn\\
\times\{<V_{\alpha}(\infty)V_{\alpha}(1)V_{p}(0)>
+z\beta_{p}^{(-1)}<V_{\alpha}(\infty)V_{\alpha}(1)L_{-1}V_{p}(0)>\nn\\
+z^{2}[\beta_{p}^{(-1,-1)}<V_{\alpha}(\infty)V_{\alpha}(1)L^{2}_{-1}V_{p}(0)>
+\beta_{p}^{(-2)}<V_{\alpha}(\infty)V_{\alpha}(1)L_{-2}V_{p}(0)>]\nn\\
+z^{3}[\beta_{p}^{(-1,-1,-1)}<V_{\alpha}(\infty)V_{\alpha}(1)L^{3}_{-1}V_{p}(0)>
+\beta_{p}^{(-1,-2)}<V_{\alpha}(\infty)V_{\alpha}(1)L_{-1}L_{-2}V_{p}(0)>\nn\\
+\beta_{p}^{(-3)}<V_{\alpha}(\infty)V_{\alpha}(1)L_{-3}V_{p}(0)>]\nn\\
+z^{4}[\beta_{p}^{(-1,-1,-1,-1)}<V_{\alpha}(\infty)V_{\alpha}(1)L^{4}_{-1}V_{p}(0)>\nn\\
+\beta_{p}^{(-1,-1-2)}<V_{\alpha}(\infty)V_{\alpha}(1)L^{2}_{-1}L_{-2}V_{p}(0)>\nn\\
+\beta_{p}^{(-1,-3)}<V_{\alpha}(\infty)V_{\alpha}(1)L_{-1}L_{-3}V_{p}(0)>
+\beta_{p}^{(-2,-2)}<V_{\alpha}(\infty)V_{\alpha}(1)L^{2}_{-2}V_{p}(0)>\nn\\
+\beta_{p}^{(-4)}<V_{\alpha}(\infty)V_{\alpha}(1)L_{-4}V_{p}(0)>]+...\}
\label{eq3.20}
\eea

We provide in (\ref{eq3.19}), (\ref{eq3.20}) the developments in powers of $z$ only. In fact, the full developments are the direct products of expansions in powers of $z$ and $\bar{z}$. But, as usual, it is sufficient to follow the $z$ development, which gives the function $F_{p}(z)$, and finally replaced it with $|F_{p}(z)|^{2}$, as in (\ref{eq3.5}), to take into account all the terms of the expansions. So that, in the developments of (\ref{eq3.19}), (\ref{eq3.20}), the developments in powers of $\bar{z}$, which are invisible, should be assumed.

The values of the matrix elements, which appear in (\ref{eq3.20}), could be calculated 
by the standard means of the conformal field theory (by moving the integrations, defining the operators $L_{-n}$, from $V_{p}(0)$ towards the other operators, etc., which is one of the methods). Their values are given in the Appendix B. Substituting these values into (\ref{eq3.20}), replacing next the series in powers of $z$, which is the conformal block function $F_{p}(z)$, by its modulus squared (as has been explained above), we obtain the decomposition (\ref{eq3.5}), with $F_{p}(z)$ given by the series:
\bea
F_{p}(z)=1+z\beta^{(-1)}_{p}\Delta_{p}
+z^{2}[\beta^{(-1,-1)}_{p}\Delta_{p}(\Delta_{p}+1)+\beta_{p}^{(-2)}(\Delta_{\alpha}+\Delta_{p})]\nn\\
+z^{3}[\beta^{(-1,-1,-1)}_{p}\Delta_{p}(\Delta_{p}+1)(\Delta_{p}+2)+\beta_{p}^{(-1,-2)}(\Delta_{\alpha}+\Delta_{p})(\Delta_{p}+2)\nn\\
+\beta_{p}^{(-3)}(2\Delta_{\alpha}+\Delta_{p})]\nn\\
+z^{4}[\beta^{(-1,-1,-1,-1)}\Delta_{p}(\Delta_{p}+1)(\Delta_{p}+2)(\Delta_{p}+3)\nn\\
+\beta^{(-1,-1,-2)}(\Delta_{\alpha}
+\Delta_{p})(\Delta_{p}+2)(\Delta_{p}+3)
+\beta_{p}^{(-1,-3)}(2\Delta_{\alpha}+\Delta_{p})(\Delta_{p}+3) \nn\\
+\beta_{p}^{(-2,-2)}(\Delta_{\alpha}+\Delta_{p})(\Delta_{\alpha}
+\Delta_{p}+2)+\beta_{p}^{(-4)}(3\Delta_{\alpha}+\Delta_{p})]+...\label{eq3.21}
\eea

We remind that $C^{p}_{\alpha,\alpha}=C_{p^{+},\alpha,\alpha}$.

The coefficients $\beta$ in the series above, or in the operator algebra expansion (\ref{eq3.19}), are defined in the standard way, [6], or lectures [11]. The system of linear equations, defining the coefficients $\beta$ up to order 4, is given in the Appendix B.

Finally we get the function $F_{p}(z)$ in the form:
\beq
F_{p}(z)=1+k^{(p)}_{1}z+k_{2}^{(p)}z^{2}+k_{3}^{(p)}z^{3}+k_{4}^{(p)}z^{4}+...\label{eq3.22}
\eeq
The values of the coefficients $k_{i}^{(p)}$, $i=1,2,3,4$ are defined 
by the expressions in (\ref{eq3.21}).

\vskip1cm

We return now to the expressions (\ref{eq3.5}), (\ref{eq3.13}) for the function \\ $<\sigma(\infty)\sigma(1)\sigma(z,\bar{z})\sigma(0)> = G(z,\bar{z})$. We shall define the principal coefficients in these formulas, $C_{\alpha,\alpha, p}  C_{p^{+},\alpha,\alpha}$ or $(D_{\alpha,\alpha, p})^{2}$, 
for the two channels available, $p_{0}$ and $p_{1}$, equations (\ref{eq3.9}), (\ref{eq3.10}).
It could be checked that, in the case of $\alpha_{\sigma}=\alpha_{\frac{3}{2},\frac{1}{2}}$, 
the calculation of the coefficients $C_{\alpha,\alpha, p}$, $C_{p^{+},\alpha,\alpha}$ 
is much simpler, compared  to the calculation of $(D_{\alpha,\alpha, p})^{2}$.

For the channel $p=p_{0}$, we get:
\beq
C_{p^{+}_{0},\alpha,\alpha}=\frac{1}{Z}\label{eq3.23}
\eeq
In fact, as $p^{+}_{0}=2\alpha_{0}-p_{0}=2\alpha_{0}-2\alpha$, we get no screenings for the 3-point function $C_{p^{+}_{0},\alpha,\alpha}$, its $l,k$ are zero. The formula (4.8) of [9] becomes trivial, gives 1, and the normalisation coefficient $1/Z$ has to be added. $Z$ is the partition function of the Coulomb Gas.

Next, it is easy to check that the 3-point function 
$C_{\alpha,\alpha, p_{0}}$ requires $l=2$, $k=0$ screenings. 
By the formula (4.8) of [9], with the factor $1/Z$ added, we obtain:
\bea
C_{\alpha\alpha p_{0}}=\frac{1}{Z}\gamma(\rho')\gamma(2\rho')\nn\\
\times\gamma^{2}(1+\alpha')\gamma^{2}(1+\alpha'+\rho')\times\gamma(1+\gamma')\gamma(1+\gamma'+\rho')\label{eq3.24}
\eea
where $\gamma(x)=\Gamma(x)/\Gamma(1-x)$, with the excuses for using the same letter 
for the function $\gamma(x)$ and the parameter $\gamma'$.
\beq
\alpha'=2\alpha_{-}\alpha=2\alpha_{-}(-\frac{\alpha_{-}}{4}+\frac{\alpha_{+}}{4})=-\frac{1+\rho'}{2}\label{eq3.25}
\eeq
$\beta'$ in (4.8), [9], is equal to $\alpha'$,
\beq
\gamma'=2\alpha_{-}p_{0}=2\alpha_{-}(-\frac{\alpha_{-}}{2}+\frac{\alpha_{+}}{2})=-(1+\rho')\label{eq3.26}
\eeq
We get, with $\rho'=2/3$,
\beq
1+\alpha'=\frac{1}{2}(1-\rho')=\frac{1}{6};\quad 1+\alpha'+\rho'=\frac{5}{6}\label{eq3.27}
\eeq
\beq
1+\gamma'=-\rho';\quad 1+\gamma'+\rho'=0\label{eq3.28}
\eeq
so that
\beq
C_{\alpha,\alpha, p_{0}}=\frac{1}{Z}\gamma(\frac{2}{3})\gamma(\frac{4}{3})\gamma^{2}(\frac{1}{6})\gamma^{2}(\frac{5}{6})\gamma(-\frac{2}{3})\gamma(0)\label{eq3.29}
\eeq 
and
\beq
C_{\alpha,\alpha, p_{0}}C_{p^{+}_{0},\alpha,\alpha}
=\frac{1}{Z^{2}}\gamma(\frac{2}{3})\gamma(\frac{4}{3})\gamma^{2}(\frac{1}{6}\gamma^{2}(\frac{5}{6})\gamma(-\frac{2}{3})\gamma(0)\label{eq3.30}
\eeq
The partition function $Z$ has been defined in [9], and it is given, 
its numerical value, in the Appendix C:
\beq
Z=-6\gamma(\frac{2}{3})\gamma(\frac{3}{2})\label{eq3.31}
\eeq
All the factors in (\ref{eq3.30}) are finite, except for $\gamma(0)=\Gamma(0)/\Gamma(1)=\infty$. So that the coefficient for the channel $p_{0}$, in the formula (\ref{eq3.5}), is infinite. But for the percolation limit point $q\rightarrow 1$, of the Potts model, this is not totally suprising. We shall comment on it shortly later.

For the channel $p=p_{1}=2\alpha+\alpha_{-}=\frac{\alpha_{-}}{2}+\frac{\alpha_{+}}{2}$ 
we shall have, for the coefficient $C_{\alpha,\alpha, p_{1}}$, $l=1$, $k=0$ 
and $C_{p^{+}_{1},\alpha,\alpha}=C_{p_{1},\alpha,\alpha}=C_{\alpha,\alpha, p_{1}}$, since
\beq
p^{+}_{1}=2\alpha_{0}-p_{1}=p_{1}\label{eq3.32}
\eeq
We shall get
\bea
C_{\alpha,\alpha, p_{1}}C_{p^{+}_{1},\alpha,\alpha}=(C_{\alpha,\alpha, p_{1}})^{2}\nn\\
=\frac{1}{Z^{2}}\gamma^{2}(\rho')\gamma^{4}(1+\alpha')\gamma^{2}(1+\gamma')\label{eq3.33}
\eea
This time
\beq
\gamma'=2\alpha_{-} p_{1}=2\alpha_{-}(\frac{\alpha_{-}}{2}+\frac{\alpha_{+}}{2})=\rho'-1, \,\,\,\, 1+\gamma' = \rho'
\label{eq3.34}
\eeq
$\alpha'$ has the same value (\ref{eq3.25}), (\ref{eq3.27}).

We obtain:
\beq
(C_{\alpha,\alpha, p_{1}})^{2}
=\frac{1}{Z^{2}} \gamma^{2}(\frac{2}{3})\gamma^{4}(\frac{1}{6})\gamma^{2}(\frac{2}{3})
\label{eq3.35}
\eeq
-- the coefficient of the channel $p_{1}$ in (\ref{eq3.5}),  is finite.

\vskip1cm

To obtain a divergent value for the coefficient of the channel $p_{0}$, not just an infinity, 
we could try to use the following regularisation:
\beq
\alpha_{+}=h\rightarrow\alpha_{+}=\tilde{h}=h+\epsilon\label{eq3.36}
\eeq
$\epsilon$ is a small regularisation parameter;  $h=\sqrt{\frac{3}{2}}$  is an unperturbed,
initial value of $\alpha_{+}$ at the percolation point.
Then
\beq
\alpha_{-}=-\frac{1}{h}\rightarrow\alpha_{-}=-\frac{1}{\tilde{h}}=-\frac{1}{h+\epsilon}
\simeq -\frac{1}{h}+\frac{2}{3}\epsilon\label{eq3.37}
\eeq
One checks that, with this regularisation, the central charge of the theory takes the value
\beq
c\simeq-\frac{10}{h}\epsilon\label{eq3.38}
\eeq
-- different from zero.

Still, it is easily seen that shifting the value of $\alpha_{+}$, as in (\ref{eq3.36}), 
and implementing the consequences which follow, is not sufficient to regularize 
the value of $C_{\alpha,\alpha, p_{0}}$, 
the factor $\gamma(1+\gamma'+\rho')$ in its expression, eq.(\ref{eq3.24}). 
This factor stays infinite. 

In addition to (\ref{eq3.36}) we shall move also the charges of the operators, as follows. 
For the 4 point function
\beq
<V_{\alpha_{4}}(\infty)V_{\alpha_{3}}(1)V_{\alpha_{2}}(z,\bar{z})V_{\alpha_{1}}(0)> \label{eq3.39}
\eeq
we shall take, instead of
\beq
\alpha_{1}=\alpha_{2}=\alpha_{3}=\alpha_{4}=\alpha=\alpha_{\frac{3}{2},\frac{1}{2}}\label{eq3.39A}
\eeq
we shall put
\beq
\alpha_{1}=\alpha+\frac{\epsilon}{2},\,\,\,\,
\alpha_{2}=\alpha+\frac{\epsilon}{2},\,\,\,\,
\alpha_{3}=\alpha-\frac{\epsilon}{2},\,\,\,\,
\alpha_{4}=\alpha-\frac{\epsilon}{2}\label{eq3.40}
\eeq
-- the choice of the shifts of the charges of the operators which is somewhat arbitrary, but such that
\beq  
\alpha_{1}+\alpha_{2}+\alpha_{3}+\alpha_{4}=4\alpha=4\alpha_{\frac{3}{2},\frac{1}{2}} 
\label{eq3.41}
\eeq
-- the total sum of the charges stays the same, unshifted, so that  the system of Coulomb Gas vertex operators could still be screened, as before, with $l=2$, $k=0$ screenings, and we could still use 
the Coulomb Gas integral representation for the correlation function (\ref{eq3.39}).

Combined regularisation, eq.(\ref{eq3.36}) and eq.(\ref{eq3.40}), was chosen so that the divergent operator algebra coefficients 
for the alternative representation of the spin operator, $\alpha_{\sigma}=\alpha_{\frac{3}{2},\frac{3}{2}}$, would also be regularized. This will be analyzed in the next Section.

In should be clear that, when the shift of the central charge of the theory,  realised by the shift of $\alpha_{+}=h$ in (\ref{eq3.36}), is followed by the shifts of the charges of operators in (\ref{eq3.40}), by doing so we are not following the Potts model line, in the parameter space of the conformal field theory. The operators with the charges (\ref{eq3.40}) are no longer the spin operators of the Potts model.

If we had shifted correctly, the charges of the operators, to stay exactly on the Potts model line, 
it would no longer be possible to screen the operators in the correlation function (\ref{eq3.39}) with integer numbers of screenings, and the integral representation could not be used in that case.

Regularising as we did, to keep valid the integral representation, we know that the finite coefficients,
finite factors in the coefficients, will keep their values, as we are keeping finally  only 
the leading approximation values, in $\epsilon$. The infinite coefficients, factors, will become divergent, in $\epsilon$, in the shift of the central charge, just as it should be the case 
for the exact continuation along the Potts model line, 
with same leading power divergence in $\epsilon$, we suppose, 
but the relative numerical values of the coefficients of these divergences, they should not be trusted, would be somewhat arbitrary, different from 
the coefficients of the divergencies in the exact Potts theory.
Qualitatively, though, the analysis will be correct. And the correlation function which remains, 
when we remove the divergent  factor in front, 
this function (of $z,\bar{z}$) will be exact, defined 
up to the overall normalisation factor.

Going back to our calculations, it could easily be checked that, with the combined regularisation of (\ref{eq3.36}), (\ref{eq3.40}), the finite factors, in the coefficient $C_{\alpha,\alpha,p_{0}}$, 
will just keep their values, in the leading order, while the infinite factor 
$\gamma(1+\gamma'+\rho')=\gamma(0)$, in equations (\ref{eq3.24}), (\ref{eq3.29}), will be replaced by
\beq
\gamma(-\frac{2\epsilon}{h})\simeq-\frac{h}{2\epsilon}\label{eq3.42}
\eeq 
In fact: $\gamma' = 2\alpha_{-}(-\frac{\alpha_{-}}{2} 
+ \frac{\alpha_{+}}{2} + \epsilon) 
= - \rho' - 1 - \frac{2}{\tilde{h}}\epsilon 
\simeq - \rho' - 1 - \frac{2}{h}\epsilon$, \\ $1+\gamma' + \rho' 
\simeq - \frac{2}{h}\epsilon$, instead of  $0$, eq.(\ref{eq3.28}), and
$\gamma(1 + \gamma' + \rho') \simeq \gamma(-\frac{2}{h}\epsilon)
\simeq - \frac{h}{2\epsilon}$.

In summary, in the expansion (\ref{eq3.5}),

\underline{$p=p_{0}$},
\beq
C_{\alpha\alpha p_{0}}C_{p^{+}_{0}\alpha\alpha}\sim\frac{1}{\epsilon}\label{eq3.43}
\eeq

\underline{$p=1$},
\beq
C_{\alpha\alpha P_{1}}C_{p^{+}_{1}\alpha\alpha}=(C_{\alpha\alpha p_{1}})^{2}, \,\,\,
\mbox{is finite}\label{eq3.44}
\eeq
If the whole function $G(z,\bar{z})=<\sigma(\infty)\sigma(1)\sigma(z,\bar{z})\sigma(0)>$ is renormalized, by multiplying it by $\epsilon$, then, in the limit $\epsilon\rightarrow 0$, only the channel $p_{0}$ will remain, and we shall get:
\beq
G(z,\bar{z})\propto\frac{1}{|z|^{4\Delta_{\alpha}-2\Delta_{p_{0}}}}\times|F_{p_{0}}(z)|^{2}\label{eq3.45}
\eeq
We remind that
\beq
\Delta_{\alpha}=\Delta_{\sigma}=\frac{5}{96},\quad\Delta_{p_{0}}=\frac{5}{8}
\label{eq3.46}
\eeq
The expansion of the conformal block function $F_{p_{0}}(z)$ is given, up to order $z^{4}$, in (\ref{eq3.21}), (\ref{eq3.22}). By substituting the numerical values of $\Delta_{\alpha}$, $\Delta_{p_{0}}$ and of the $\beta$-coefficients, which are defined in the Appendix B, we get the following values
of the coefficients in (\ref{eq3.22}): 
\bea
k_{1} = \frac{5}{16}, \quad k_{2} = \frac{845}{4608}, \quad
k_{3} = \frac{3185}{24576}, \quad k_{4} = \frac{4257635}{42467328}\nn\\
k_{1} \simeq 0.3125, \quad k_{2} \simeq 0.1834, 
\quad k_{3} \simeq 0.1296, \quad k_{4} \simeq 0.1003 \label{eq3.47A}
\eea
and the following expansion for the function $F_{p_{0}}(z)$:
\beq
F_{p_{0}}(z)=1+0.3125 z+0.1834 z^{2}+0.1296 z^{3}+0.1003 z^{4}+...\label{eq3.47B}
\eeq

\vskip1cm

Above we have calculated the coefficients $C_{\alpha,\alpha, p}C_{p^{+},\alpha,\alpha}$ in the formula (\ref{eq3.5}). Alternatively we could have used the expansion in (\ref{eq3.13}), with coefficients $(D_{\alpha,\alpha, p})^{2}$.
Calculations, using the formula (4.41) of [9], would be more complicated, in the present case 
of $\alpha=\alpha_{\frac{3}{2},\frac{1}{2}}$.  While in the case of $\alpha=\alpha_{\frac{3}{2},\frac{3}{2}}$, which will be analysed in the next Section, it will be just the opposite, calculation of the coefficient
$(D_{\alpha,\alpha, p})^{2}$ will be much simpler.

The relation between the coefficients $C_{\alpha,\alpha, p}C_{p^{+},\alpha,\alpha}$ 
and $(D_{\alpha,\alpha, p})^{2}$ is given by the formula (\ref{eq3.15}). 
The numerical value of the coefficient $(N_{\alpha})^{4}/Z$, for $\alpha=\alpha_{\frac{3}{2},\frac{1}{2}}$, is defined in the Appendix C:
\beq
\frac{(N_{\alpha_{\frac{3}{2},\frac{1}{2}}})^{4}}{Z}=-\frac{1}{486}\cdot\frac{1}{\gamma^{2}(\frac{3}{4})\gamma^{3}(\frac{3}{2})\gamma^{3}(\frac{2}{3})}\label{eq3.48}
\eeq
For our present calculations, the only importance is that this coefficient contains no singularities, it is finite. So that, similarly, $(D_{\alpha,\alpha, p_{0}})^{2}$, in the expansion (\ref{eq3.13}),  is going to be divergent, 
as $\frac{1}{\epsilon}$, and $(D_{\alpha,\alpha, p_{1}})^{2}$ will be finite, This have been checked, actually, by the direct calculation of the coefficients 
$D_{\alpha,\alpha, p_{0}}$, $D_{\alpha,\alpha, p_{1}}$, and the formula (\ref{eq3.15}) has been verified.

With respect to the divergence of one of the coefficients in the expansion (\ref{eq3.5}), this could be compared with the analysis in [2] of the 4-spin function of the Potts model, as defined by the cluster expansion on the lattice. In particular, the equation (19) of [2] is of the form:
\bea
G_{\alpha\alpha\alpha\alpha}=(q-1)(q^{2}-3q+3)P_{aaaa}\nn\\
+(q-1)^{2}(P_{aabb}+P_{abba}+P_{abab})\label{eq3.49}
\eea
$G_{\alpha\alpha\alpha\alpha}$ is the 4-spin correlation function, while $P_{aaaa}$, $P_{aabb}$, etc., are the cluster connectivities, which are finite, involve no singularities as $q\rightarrow 1$.

For the 2-spin function, eq.(16) of [2], one has:
\beq
G_{\alpha\alpha}=(q-1)P_{aa}\label{eq3.50}
\eeq
Two-point connectivity $P_{aa}$ is finite, as $q\rightarrow 1$. If we renormalize, spins, so that the two-point function in (\ref{eq3.50}) becomes finite, as $q\rightarrow 1$,
\beq
\tilde{G}_{\alpha\alpha}=P_{aa}\label{eq3.51}
\eeq
then the formula for the 4-spin function in (\ref{eq3.49}) will take the form, for $q$ close to 1 
$(q^{2}-3q+3 \simeq 1)$:
\beq
\tilde{G}_{\alpha\alpha\alpha\alpha} \simeq \frac{1}{q-1}P_{aaaa}+(P_{aabb}+P_{abba}+P_{abab})\label{eq3.52}
\eeq
It is similar to our expansion in (\ref{eq3.5}), one of the coefficients is divergent, 
in the limit $q\rightarrow 1$. Similarly to the normalisation above, having the two point function
finite as $q \rightarrow 1$,  in our calculations the two-point function is also finite : 
the norm squared, $(N_{\alpha})^{2}$,  of the spin operator, is finite, Appendix C.

Still, it should be noted that (\ref{eq3.5}) and (\ref{eq3.52}) are not the same. The representations for the spin operators in [2], and in our formulas, are different. In [2], in the function $G_{\alpha\alpha\alpha\alpha}$, the spins are fixed to have a definite value $\alpha$, out of $q$ possible values. Like, in the case of $Z_{4}$ symetric spins (just as an exemple), taking 4 values, we would fixe the spins in the direction 1 and calculate the correlator $<\sigma^{1}\sigma^{1}\sigma^{1}\sigma^{1}>$.

In the conformal theory, on the other hand, one is working with the $Z_{4}$ Fourier components, 
having fixed values of $Z_{4}$ spins: let us note them as $\sigma_{1}$, the first $Z_{4}$ Fourier
component,  
the operator having the $Z_{4}$ spin 1, $\sigma_{2}$, the operator having the $Z_{4}$ spin 2, and next the operators $\sigma_{-2}$ and $\sigma_{-1}$. One is working with the operators having also definite conformal dimensions. They have the fusions
\beq
\sigma_{1}\times \sigma_{1}\rightarrow\sigma_{2}\label{eq3.53}
\eeq
etc. . In particular, $\sigma_{1}$ is not reproduced when fusing $\sigma_{1}\times\sigma_{1}$.

It is different when working with spins having definite orientations, $\sigma^{1},\sigma^{2},\sigma^{3},\sigma^{4}$. In particular, $\sigma^{1}$ could be expressed as a linear combination of all the operators $\sigma_{1},\sigma_{2},\sigma_{-2},\sigma_{-1}$, having definite $Z_{4}$ spin. When fusing $\sigma^{1}\times\sigma^{1}$, naturally, the operator $\sigma^{1}$ will be reproduced, among the others.

In this argument we have used $Z_{4}$ just as an example, to stress the difference between the two different representations of spin operators, leading to different fusion rules.

Coming back to the formula (19) of [2], reproduced here as (\ref{eq3.49}), (\ref{eq3.52}), 
the function $G_{\alpha\alpha\alpha\alpha}$ will have, naturally, the spin operator 
among its intermediate channels, with dimension $\Delta=5/96$. 

En the other hand, for the conformal theory spins, we have found, 
for $<\sigma\sigma\sigma\sigma>$, 
a single intermediate channel, having the dimension $\Delta_{p_{0}}=5/8$. 
We would suggest that this channel is present also
 in the function $G_{\alpha\alpha\alpha\alpha}$, but as a sub-leading one. 
 Like this would be the case with two types of four-spin functions
  in the example of the model of $Z_{4}$ invariant spins.
 
In the next Section we shall check the consistency 
on the conformal theory side, 
that the calculations, with the spin operator represented 
by the vertex operator having  the charge 
$\alpha_{\sigma}=\alpha_{\frac{3}{2},\frac{3}{2}}$, lead to the same four-point function, the one in (\ref{eq3.45}).

\section{Consistency check: calculation of the function \\ $<\sigma \sigma \sigma \sigma>$
with the representation $\alpha_{\frac{3}{2},\frac{3}{2}}$  for the spin operator.}

In the case when the spin operator is represented by the  vertex operator
\beq
V_{\alpha},\,\,\,\alpha=\alpha_{\frac{3}{2},\frac{3}{2}}=-\frac{\alpha_{-}}{4}-\frac{\alpha_{+}}{4}\label{eq4.1}
\eeq 
for the integral representation (\ref{eq2.5}) of the function
\beq
<\sigma\sigma\sigma\sigma>\,\,\propto
 \,\, <V_{\alpha}V_{\alpha}V_{\alpha}V_{\alpha}>_{conf}
\label{eq4.2}
\eeq
on finds, by the neutrality condition (\ref{eq2.7}), that one needs
\beq
l=2,\quad k=2\label{eq4.3}
\eeq
screenings $V_{-}$ and $V_{+}$, in (\ref{eq2.5}). So that, instead of the double $2D$ integral in (\ref{eq3.4}), for the case of $\alpha_{\sigma}=\alpha_{\frac{3}{2},\frac{1}{2}}$, one will obtain the integral
\bea
\int d^{2}u_{1}\int d^{2}u_{2}\int d^{2}v_{1}\int d^{2}v_{2}\,
<V_{\alpha}(\infty)V_{\alpha}(1)V_{\alpha}(z,\bar{z})V_{\alpha}(0)\cdot V_{-}(u_{1},\bar{u}_{1})V_{-}(u_{2},\bar{u}_{2})\nn\\
V_{+}(v_{1},\bar{v}_{1})V_{+}(v_{2},\bar{v}_{2})>\label{eq4.4}
\eea
for the function
\beq
<\sigma(\infty)\sigma(1)\sigma(z,\bar{z})\sigma(0)> \,\,
\propto \,\, <V_{\alpha}(\infty)V_{\alpha}(1)V_{\alpha}(z,\bar{z})V_{\alpha}(0)>_{conf.}\label{eq4.5}
\eeq
By factorizing this integral, as we did in the Section 3 for the integral (\ref{eq3.4}), we shall find, in the present case,
\bea
G'(z,\bar{z}) \equiv \,\, <V_{\alpha}(\infty)V_{\alpha}(1)V_{\alpha}(z,\bar{z})V_{\alpha}(0)>\nn\\
\propto \sum_{i}\frac{1}{|z|^{4\Delta_{\alpha}-2\Delta_{p'_{i}}}}
\,C_{\alpha,\alpha,p'_{i}}C_{p'^{+}_{i},\alpha,\alpha}\,
|F_{p'_{i}}(z)I^{2}\label{eq4.6}
\eea
or
\beq
G'(z,\bar{z})\propto\sum_{i}
\frac{1}{|z|^{4\Delta_{\alpha}-2\Delta_{p'_{i}}}}
\,(D_{\alpha,\alpha,p'_{i}})^{2}\, |F_{p'_{i}}(z)|^{2}\label{eq4.7}
\eeq
In the above $\alpha$ is given in (\ref{eq4.1}). We denote with primes $G'(z,\bar{z})$, $p'_{i}$, the function and the parameter $p$, to make the difference with the corresponding quantities which have been defined and analysed in the previous Section, for the representation $\alpha=\alpha_{\frac{3}{2},\frac{1}{2}}$  of the spin operator.

The intermediate channels in (\ref{eq4.6}) or (\ref{eq4.7}) have the following charges $\{p'_{i}\}$:
\bea
p'_{0}=2\alpha=-\frac{\alpha_{-}}{2}-\frac{\alpha_{+}}{2}\nn\\
p'_{1}=2\alpha+\alpha_{-}=\frac{\alpha_{-}}{2}-\frac{\alpha_{+}}{2}\nn\\
p'_{2}=2\alpha+\alpha_{+}=-\frac{\alpha_{-}}{2}+\frac{\alpha_{+}}{2}\nn\\
p'_{3}=2\alpha+2\alpha_{-}=\frac{3}{2}\alpha_{-}-\frac{\alpha_{+}}{2}\nn\\
p'_{4}=2\alpha+\alpha_{-}+\alpha_{+}=\frac{\alpha_{-}}{2}+\frac{\alpha_{+}}{2}\nn\\
p'_{5}=2\alpha+2\alpha_{+}=-\frac{\alpha_{-}}{2}+\frac{3}{2}\alpha_{+}\nn\\
p'_{6}=2\alpha+2\alpha_{-}+\alpha_{+}=\frac{3}{2}\alpha_{-}+\frac{\alpha_{+}}{2}\nn\\
p'_{7}=2\alpha+\alpha_{-}+2\alpha_{+}=\frac{\alpha_{-}}{2}+\frac{3}{2}\alpha_{+}\nn\\
p'_{8}=2\alpha+2\alpha_{-}+2\alpha_{+}=\frac{3}{2}\alpha_{-}+\frac{3}{2}\alpha_{+}\label{eq4.8}
\eea
But one can check that
\bea
p'^{+}_{0}=2\alpha_{0}-p'_{0}=p'_{8},\,\,
p^{+}_{1}=2\alpha_{0}-p'_{1}=p'_{7},\,\,
p'^{+}_{2}=2\alpha_{0}-p'_{2}=p'_{6}\nn\\
p'^{+}_{3}=2\alpha_{0}-p'_{3}=p'_{5},\,\,p'^{+}_{4}=2\alpha_{0}-p'_{4}=p'_{4}\label{eq4.9}
\eea
so that the channels 8,7,6,5 are equivalent to the channels 0,1,2,3,  the corresponding terms in (\ref{eq4.6}) or (\ref{eq4.7}) are equal between themselves.

We find that there are 5 independent channels
\beq
p'_{0}, \, p'_{1}, \,p'_{2}, \,p'_{3}, \,p'_{4}\label{eq4.10}
\eeq
in the expansions (\ref{eq4.6}), (\ref{eq4.7}), instead of 9 listed in (\ref{eq4.8}).

We observe also that the channels $p'_{2}$ and $p'_{4}$ correspondent, respectively, to the channels $p_{0}$ and $p_{1}$ of the previous Section. Also, because, eq.(\ref{eqC.14}),
\beq
\alpha_{\frac{3}{2},\frac{3}{2}}=\alpha^{+}_{\frac{3}{2},\frac{1}{2}}=2\alpha_{0}-\alpha_{\frac{3}{2},\frac{1}{2}}\label{eq4.10A}
\eeq
the channels $p'_{2}$ and $p'_{4}$ don't have 
to be analysed, their contributions to the expansion (\ref{eq4.6}) or (\ref{eq4.7}) 
are equal to those calculated in the previous Section, for the channels $p_{0}$ and $p_{1}$. In fact, for instance:
\beq
D_{\alpha_{\frac{3}{2},\frac{3}{2}},\alpha_{\frac{3}{2},\frac{3}{2}},p'_{2}}
=D_{\alpha^{+}_{\frac{3}{2},\frac{1}{2}},\alpha^{+}_{\frac{3}{2},
\frac{1}{2}},p_{0}}
=D_{\alpha_{\frac{3}{2},\frac{1}{2}},
\alpha_{\frac{3}{2},\frac{1}{2}},p_{0}}\label{eq4.11}
\eeq
We have used here the fact that the coefficients $D_{\alpha,\alpha,p}$ are symmetric with respect to the conjugation, of any of its indices. Also we remind that $\Delta_{\alpha_{\frac{3}{2},\frac{3}{2}}}=\Delta_{\alpha_{\frac{3}{2},\frac{1}{2}}}$.

There remain the channels $p'_{0}$, $p'_{1}$ and $p'_{3}$, in the liste (\ref{eq4.8}), to be analysed. In fact, to obtain the same function 
$G(z,\bar{z})$ as in the previous Section, i.e. to obtain that 
$G'(z,\bar{z})=G(z,\bar{z})$, we have to show that the contribution of the channels $p'_{0}$, $p'_{1}$, and $p'_{3}$, to the sum in (\ref{eq4.6}) or in (\ref{eq4.7}), their contribution just vanish.

In the following we shall calculate and analyse the contributions of these three channels. And first we shall do the calculations, of the corresponding coefficients $(D_{\alpha,\alpha,p})^{2}$ in (\ref{eq4.7}), without any regularisation, similarly as we did it in the previous Section, for the channels $p_{0}$ and $p_{1}$.
We remind that the relation between the coefficients $C_{\alpha,\alpha,p}C_{p^{+},\alpha,\alpha}$ and $(D_{\alpha,\alpha,p})^{2}$, in (\ref{eq4.6}) and (\ref{eq4.7}), is given by the eq.(\ref{eq3.15}). Compared to the Section 3, the value of $\alpha$ has changed. For $\alpha=\alpha_{\frac{3}{2},\frac{3}{2}}$ we get, Appendix C,
\beq
\frac{(N_{\alpha})^{4}}{Z}=-\frac{1}{96}\cdot\frac{\gamma^{2}(\frac{3}{4})}{\gamma^{3}(\frac{3}{2})\gamma^{3}(\frac{2}{3})}\label{eq4.12}
\eeq
-- instead of (\ref{eq3.48}). This coefficient is finite. According to the Section 3, we are principally concerned by the coefficients $C_{\alpha,\alpha,p}C_{p^{+},\alpha,\alpha}$ or $(D_{\alpha,\alpha,p})^{2}$ which are divergent. After rescaling the correlation function, i.e. after multiplying the function $<\sigma\sigma\sigma\sigma>$ by $\epsilon$, the parameter of the regularization, and taking the limite $\epsilon \rightarrow 0$, 
only the channels with the coefficients $C_{\alpha,\alpha,p}C_{p^{+},\alpha,\alpha}$ or $(D_{\alpha,\alpha,p})^{2}$ which are divergent will remain. So we can calculate $C_{\alpha,\alpha,p}C_{p^{+},\alpha,\alpha}$ or $(D_{\alpha,\alpha,p})^{2}$, whichever is simpler, looking, principally, for divergences.

\underline{For $p'_{0}$}, the calculation of $(D_{\alpha,\alpha,p'_{0}})^{2}$ is simpler. We find:
\beq
\alpha=\alpha_{\frac{3}{2},\frac{3}{2}}=-\frac{\alpha_{-}}{4}-\frac{\alpha_{+}}{4},\,\,p'_{0}=2\alpha=-\frac{\alpha_{-}}{2}-\frac{\alpha_{+}}{2}=-\alpha_{0}\label{eq4.13}
\eeq

By the formula (4.41), [9], we obtain:
\bea
(D_{\alpha,\alpha,p'_{0}})^{2}=\frac{\Upsilon^{2}(2\alpha+p'_{0}-2\alpha_{0})\Upsilon^{4}(p'_{0})\Upsilon(2\alpha-p'_{0})\Upsilon(-2\alpha_{0})}
{\Upsilon^{2}(2\alpha)\Upsilon^{2}(2\alpha-2\alpha_{0})\Upsilon(2p'_{0})\Upsilon(2 p'_{0}-2\alpha_{0})}\nn\\
=\frac{\Upsilon^{2}(-4\alpha_{0})\Upsilon^{4}(-\alpha_{0})\Upsilon^{2}(0)\Upsilon(-2\alpha_{0})}
{\Upsilon^{2}(-\alpha_{0})\Upsilon^{2}(-3\alpha_{0})\Upsilon(-2\alpha_{0})\Upsilon(-4\alpha_{0})}\nn\\
=\frac{\Upsilon(-4\alpha_{0})\Upsilon^{2}(-\alpha_{0})}
{\Upsilon^{2}(-3\alpha_{0})}\label{eq4.14}
\eea
We have used the value $\Upsilon(0)=1$, [9], Appendix B.
According to the Appendix C, of the present paper,
\beq
\frac{\Upsilon(-\alpha_{0})}{\Upsilon(-3\alpha_{0})}=\frac{h^{1/2}}{\gamma(\frac{3}{4})}\label{eq4.15}
\eeq
We remind that $h \equiv \alpha_{+}=\sqrt{3/2}$. We obtain
\beq
(D_{\alpha,\alpha,p'_{0}})^{2}=\Upsilon(-4\alpha_{0})\cdot\frac{h}{\gamma^{2}(\frac{3}{4})}\label{eq4.16}
\eeq
Next, 
\bea
2\alpha_{0}=\alpha_{-}+\alpha_{+}=-\frac{1}{h}+h\nn\\
=\frac{1}{h}(-1+h^{2})=\frac{1}{h}\cdot\frac{1}{2};\nn\\
\alpha_{0}=\frac{1}{4h},\quad 4\alpha_{0}=\frac{1}{h}\label{eq4.17}
\eea
We get
\beq
(D_{\alpha,\alpha,p'_{0}})^{2}=\Upsilon(-\frac{1}{h})\cdot\frac{h}{\gamma^{2}(\frac{3}{4})}\label{eq4.18}
\eeq
But $\Upsilon(-\frac{1}{h})=0$, Appendix B of [9]. So that
\beq
(D_{\alpha,\alpha,p'_{0}})^{2}=0\label{eq4.19}
\eeq

We remind that we are doing the first calculation of the coefficients in (\ref{eq4.6}) or (\ref{eq4.7}) without regularisation.

Now, \underline{the channel $p'_{1}$.}
\bea
\alpha=-\frac{\alpha_{-}}{4}-\frac{\alpha_{+}}{4},\,\,2\alpha=-\frac{\alpha_{-}}{2}-\frac{\alpha_{+}}{2}=-\alpha_{0},\,\,4\alpha=-2\alpha_{0}\nn\\
p'_{1}=2\alpha+\alpha_{-}=\frac{\alpha_{-}}{2}-\frac{\alpha_{+}}{2}=-\alpha_{0}+\alpha_{-}\nn\\
\alpha_{0}=\frac{1}{4h},\,\,\alpha_{-}=-\frac{1}{h}\label{eq4.20}
\eea
\bea
(D_{\alpha,\alpha,p_{1}})^{2}=\frac{\Upsilon^{2}(2\alpha+p'_{1}-2\alpha_{0})\Upsilon^{4}(p'_{1})\Upsilon^{2}(2\alpha-p'_{1})\Upsilon(-2\alpha_{0})}
{\Upsilon^{2}(2\alpha)\Upsilon^{2}(2\alpha-2\alpha_{0})\Upsilon(2p'_{1})\Upsilon(2p'_{1}-2\alpha_{0})}\nn\\
=\frac{\Upsilon^{2}(4\alpha+\alpha_{-}-2\alpha_{0})\Upsilon^{4}(2\alpha+\alpha_{-})\Upsilon^{2}(-\alpha_{-})\Upsilon(-2\alpha_{0})}
{\Upsilon^{2}(-\alpha_{0})\Upsilon^{2}(-3\alpha_{0})\Upsilon(4\alpha+2\alpha_{-})\Upsilon(4\alpha+2\alpha_{-}-2\alpha_{0})}\nn\\
=\frac{\Upsilon^{2}(-4\alpha_{0}+\alpha_{-})\Upsilon^{4}(-\alpha_{0}+\alpha_{-})\Upsilon^{2}(-\alpha_{-})\Upsilon(-2\alpha_{0})}
{\Upsilon^{2}(-\frac{1}{4h})\Upsilon^{2}(-\frac{3}{4h})\Upsilon(-2\alpha_{0}+2\alpha_{-})\Upsilon(-4\alpha_{0}+2\alpha_{-})}\label{eq4.21}
\eea
Here $\Upsilon^{2}(-\frac{1}{4h})=\Upsilon^{2}(-\alpha_{0})$, $\Upsilon^{2}(-\frac{3}{4h})=\Upsilon^{2}(-3\alpha_{0})$, $\Upsilon(-2\alpha_{0})$, they are all finite, and non-zero, Appendix C. Also $\Upsilon^{2}(-\alpha_{-})=\Upsilon^{2}(\frac{1}{h})=\Upsilon^{2}(4\alpha_{0})=\Upsilon^{2}(-2\alpha_{0})$ is finite. In the last equality we have use the property of the $\Upsilon$ function $\Upsilon(x)=\Upsilon(2\alpha_{0}-x)$.

So that, from (\ref{eq4.21}), we obtain:
\bea
(D_{\alpha,\alpha,p'_{1}})^{2}\propto\frac{\Upsilon^{2}(-4\alpha_{0}+\alpha_{-})\Upsilon^{4}(-\alpha_{0}+\alpha_{-})}
{\Upsilon(-2\alpha_{0}+2\alpha_{-})\Upsilon(-4\alpha_{0}+2\alpha_{-})}\nn\\
=\frac{\Upsilon^{2}(-\frac{2}{h})\Upsilon^{4}(-\frac{5}{4h})}{\Upsilon(-\frac{5}{2h})\Upsilon(-\frac{3}{h})}\label{eq4.22}
\eea 
$\Upsilon^{4}(-\frac{5}{4h})$ is finite. In fact, by the property, Appendix B of [9]:
\beq
\Upsilon(x-\frac{1}{h})=\frac{1}{\gamma(\frac{x}{h})}h^{1-\frac{2x}{h}}\times\Upsilon(x)\label{eq4.23}
\eeq
we obtain
\bea
\Upsilon(-\frac{5}{4h})=\Upsilon(-\frac{1}{4h} - \frac{1}{h})
=\frac{1}{\gamma(-\frac{1}{4h^{2}})}h^{1+\frac{2}{h}\cdot\frac{1}{4h}}\times\Upsilon(-\frac{1}{4h})\nn\\
=\frac{1}{\gamma(-\frac{1}{6})}h^{4/3}\times\Upsilon(-\frac{1}{4h})\label{eq4.24}
\eea
As $\Upsilon(-\frac{1}{4h})$ is finite, Appendix C, $\Upsilon(-\frac{5}{4h})$ is also finite and non-zero.

 Also, by (\ref{eq4.23}),
\bea
\Upsilon(-\frac{3}{h})= \Upsilon(-\frac{2}{h} - \frac{1}{h}) =
\frac{1}{\gamma(-\frac{2}{h^{2}})}h^{1+\frac{4}{h^{2}}}\times\Upsilon(-\frac{2}{h})\nn\\
=\frac{1}{\gamma(-\frac{4}{3})}h^{11/3}\times\Upsilon(-\frac{2}{h})\label{eq4.25}
\eea
Taking into account (\ref{eq4.24}), (\ref{eq4.25}), eq.(\ref{eq4.22}) takes the form:
\beq
(D_{\alpha,\alpha,p'_{1}})^{2}\propto\frac{\Upsilon(-\frac{2}{h})}{\Upsilon(-\frac{5}{2h})}\label{eq4.26}
\eeq
$\Upsilon(-\frac{5}{2h})$ could be transformed as follows:
\bea
-\frac{5}{2h}=-\frac{5}{2h^{2}}\cdot h=-\frac{5}{3}h\nn\\
\Upsilon(-\frac{5}{2h})=\Upsilon(-\frac{5}{3}h)=\Upsilon(-\frac{2}{3}h-h)=\Upsilon(-\frac{1}{h}-h)\nn\\
=\gamma(-h(-\frac{5}{3}h))h^{1+2h\cdot(\frac{-5}{3}h)}\times\Upsilon(-\frac{1}{h})\nn\\
=\gamma(\frac{5}{2})h^{-4}\times\Upsilon(-\frac{1}{h})\label{eq4.27}
\eea
We have used the property, Appendix B of [9],
\beq
\Upsilon(x-h)=\gamma(-h (x-h))h^{1+2h (x-h)}\times\Upsilon(x)\label{eq4.28}
\eeq
Also for $\Upsilon(-\frac{2}{h})$ in {\ref{eq4.26}), we obtain, by (\ref{eq4.23}):
\bea
\Upsilon(-\frac{2}{h}) =  \Upsilon(-\frac{1}{h} - \frac{1}{h}) =
\frac{1}{\gamma(-\frac{1}{h^{2}})}h^{1+\frac{2}{h^{2}}}\times\Upsilon(-\frac{1}{h})\nn\\
=\frac{1}{\gamma(-\frac{2}{3})}h^{7/3}\times\Upsilon(-\frac{1}{h})\label{eq4.29}
\eea
Taking into account (\ref{eq4.29}) and (\ref{eq4.27}), for $\Upsilon(-\frac{2}{h})$ and $\Upsilon(-\frac{5}{2h})$ in (\ref{eq4.26}), we can conclude that $(D_{\alpha,\alpha,p'_{1}})^{2}$ is finite.

\underline{Conclusion:}

Under rescaling $\epsilon <\sigma\sigma\sigma\sigma>$, $\epsilon\rightarrow 0$, the channel $p'_{1}$ will disappear.

\newpage

\underline{The channel $p'_{3}$}

For this channel we shall calculate $C_{\alpha,\alpha,p'_{3}}C_{p'^{+}_{3},\alpha,\alpha}$, which is simpler, instead of $(D_{\alpha,\alpha,p'_{3}})^{2}$.
\bea
p'_{3}=2\alpha+2\alpha_{-}=-\alpha_{0}+2\alpha_{-}=\frac{3}{2}\alpha_{-}-\frac{1}{2}\alpha_{+}\nn\\
p'^{+}_{3}=2\alpha_{0} - p'_{3} = 2\alpha_{0} - 2\alpha - 2\alpha_{-}
= 3\alpha_{0} - 2\alpha_{-} 
=-\frac{1}{2}\alpha_{-}+\frac{3}{2}\alpha_{+} = 2\alpha + 2\alpha_{+}
\label{eq4.30}
\eea

\underline{$C_{\alpha,\alpha,p'_{3}}$.}

For the Coulomb Gas formula, eq.(\ref{eq4.8}) of [9] (with $\frac{1}{Z}$ to be added), for the three-point function $C_{\alpha,\alpha,p'_{3}}$, one needs {$l=0, k=2$} screenings.

In fact:
\bea
2\alpha+p'_{3}+l\alpha_{-}+k\alpha_{+}=2\alpha_{0}\nn\\
4\alpha+2\alpha_{-}+l\alpha+k\alpha_{+}=\alpha_{-}+\alpha_{+}\nn\\
-\alpha_{-}-\alpha_{+}+2\alpha_{-}+l\alpha_{-}+k\alpha_{+}=\alpha_{-}+\alpha_{+}\nn\\
l\alpha_{-}+k\alpha_{+}=2\alpha_{+}\rightarrow l=0, k=2 \label{eq4.31}
\eea
The parameters in the formula (4.8) of [9]:
\bea
\alpha=2\alpha_{+}\alpha_{\sigma}=2\alpha_{+}(-\frac{\alpha_{-}}{4}-\frac{\alpha_{+}}{4})=\frac{1-\rho}{2}=-\frac{1}{4}\nn\\
\gamma=2\alpha_{+}p'_{3}=2\alpha_{+}(\frac{3}{2}\alpha_{-}-\frac{1}{2}\alpha_{+})=-3-\rho=-\frac{9}{2}\label{eq4.32}
\eea
There is a small interference between the parameter $\alpha$ of the Coulomb Gas formula and the charge $\alpha$ of the spin operator, which we denoted here more explicitly as $\alpha_{\sigma}$.

By the formula (4.8) of [9] we find;
\bea
C_{\alpha,\alpha,p'_{3}}=\frac{1}{Z}\gamma(\rho)\gamma(2\rho)\nn\\
\times\gamma^{2}(1+\alpha)\gamma^{2}(1+\alpha+\rho)\times\gamma(1+\gamma)\cdot\gamma(1+\gamma+\rho)\label{eq4.33}
\eea
Another interference: the function $\gamma(x)=\Gamma(x)/\Gamma(1-x)$ and the parameter $\gamma$.

We obtain
\bea
C_{\alpha,\alpha,p'_{3}}=\frac{1}{Z}\times\gamma(\frac{3}{2})\gamma(3)\nn\\
\times\gamma^{2}(\frac{3}{4})\gamma^{2}(\frac{9}{4})\times\gamma(-\frac{7}{2})\gamma(-2)\label{eq4.34}
\eea
In the above, the two factors are singular:
\bea
\gamma(3)=\frac{\Gamma(3)}{\Gamma(-2)}\sim\frac{1}{\Gamma(-2)}\nn\\
\gamma(-2)=\frac{\Gamma(-2)}{\Gamma(3)}\sim\Gamma(-2)\label{eq4.35}
\eea
One could suppose that they simplify one another, but this will have to be made more precise with the regularised calculation. At least, it is not obvions, at all, that their product is equal to 1. But we could suppose, naturally, that
\beq
\gamma(3) \propto \epsilon ,\quad \gamma(-2) \propto \frac{1}{\epsilon}
\label{eq4.36}
\eeq
and that $C_{\alpha,\alpha,p'_{3}}$ is finite.

\underline{$C_{p'^{+}_{3},\alpha,\alpha}$}

For this 3-point function one needs $l=2, k=0$ screenings. The relevant parameters:
\bea
\alpha'=2\alpha_{-}\alpha_{\sigma}=2\alpha_{-}(-\frac{\alpha_{-}}{4}-\frac{\alpha_{+}}{4})=\frac{1-\rho'}{2}=\frac{1}{6}\nn\\
\gamma'=2\alpha_{-} p'^{+}_{3}=2\alpha_{-}(-\frac{1}{2}\alpha_{-}+\frac{3}{2}\alpha_{+})=-3-\rho'=-\frac{11}{3}\label{eq4.37}
\eea
With the formula (4.8), [9], we obtain:
\bea
C_{p'^{+}_{3},\alpha,\alpha}=\frac{1}{Z}\gamma(\rho')\gamma(2\rho')
\times\gamma^{2}(1+\alpha')\gamma^{2}(1+\alpha'+\rho')\times\gamma(1+\gamma')\gamma(1+\gamma'+\rho')\nn\\
=\frac{1}{Z}\gamma(\frac{2}{3})\gamma(\frac{4}{3})\times\gamma^{2}(\frac{7}{6})\gamma^{2}(\frac{11}{6})\times\gamma(-\frac{8}{3})\gamma(-2)\label{eq4.38}
\eea
There is one divergente factor, $\gamma(-2)\sim\Gamma(-2)$. As a result, $C_{p'^{+}_{3},\alpha,\alpha}$ is divergente, and the coefficient $C_{\alpha,\alpha,p'_{3}}\cdot C_{p'_{3},\alpha,\alpha}$, in the expansion (\ref{eq4.6}) 
of the function $<\sigma\sigma\sigma\sigma>$, is divergent. It appears that, after the rescaling $\epsilon <\sigma\sigma\sigma\sigma>$, 
$\epsilon \rightarrow 0$, the channel $p'_{3}$ will contribute, so that the 4-spin function, calculated with the representation $\alpha_{\sigma}=\alpha_{\frac{3}{2},\frac{3}{2}}$ would be different compared to the function which we calculated on the Section 3, with $\alpha_{\sigma}=\alpha_{\frac{3}{2},\frac{1}{2}}$.

But let us look more closely at the probleme.

The conformal dimensions of 5 channels (\ref{eq4.10}) have the following values:
\bea
\Delta_{p'_{0}}=\frac{1}{8},\quad\Delta_{p'_{1}}=\frac{35}{24},\quad\Delta_{p'_{2}}=\frac{5}{8},\nn\\
\Delta_{p'_{3}}=\frac{33}{8},\quad \Delta_{p'_{4}}=-\frac{1}{24}\label{eq4.39}
\eea
We observe that there is interference, on resonance, between the channels $p'_{0}$ and $p'_{3}$:
\beq
\Delta_{p'_{3}}-\Delta_{p'_{0}}=4\label{eq4.40}
\eeq
There are no others resonances in between the 5 channels above.

The channel $p'_{0}$, according to our estimations, should have the coefficient
\beq
(D_{\alpha,\alpha,p'_{0}})^{2}\sim\epsilon\label{eq4.41}
\eeq
-- after regularisation, compare (\ref{eq4.19}), because only one factor, in the expression for $(D_{\alpha,\alpha,p'_{0}})^{2}$, is vanishing, and because 
all $\epsilon$ corrections, to the expressions for the parameters, should be linear in $\epsilon$.

We shall verify our estimations with the regularised calculations below. 

The coefficient $C_{\alpha,\alpha,p'_{3}}C_{p'^{+}_{3},\alpha,\alpha}$, or, equivalently, the coefficient $(D_{\alpha,\alpha,p'_{3}})^{2}$, of the channel $p'_{3}$, in the expansion (\ref{eq4.7}), should behave as:
\beq
(D_{\alpha,\alpha,p'_{3}})^{2}\sim\frac{1}{\epsilon}\label{eq4.42}
\eeq
-- according to our estimations, compare the expression in eq.(\ref{eq4.38}) which has only one divergent factor, $\gamma(-2)$.

Now, looking at the contribution of the two channels above to the expansion (\ref{eq4.7}):
\bea
\frac{(D_{\alpha,\alpha,p_{0}})^{2}}{|z|^{4\Delta_{\alpha}-2\Delta_{p'_{0}}}}|F_{p'_{0}}(z)|^{2}\nn\\
+\frac{(D_{\alpha,\alpha,p'_{3}})^{2}}{|z|^{4\Delta_{\alpha}-2\Delta_{p'_{3}}}}|F_{p'_{3}}(z)|^{2}\nn\\
=\frac{1}{|z|^{4\Delta_{\alpha}-2\Delta_{p'_{0}}}}\times\{(D_{\alpha,\alpha,p'_{0}})^{2}|F_{p'_{0}}(z)|^{2}\nn\\
+(D_{\alpha,\alpha,p'_{3}})^{2}|z|^{2(\Delta_{p'_{3}}-\Delta_{p'_{0}})}|F_{p'_{3}}(z)|^{2}\}\nn\\
=\frac{1}{|z|^{4\Delta_{\alpha}-2\Delta_{p'_{0}}}}\times\{(D_{\alpha,\alpha,p'_{0}})^{2}|F_{p'_{0}},(z)|^{2}\nn\\
+(D_{\alpha,\alpha,p'_{3}})^{2}|z|^{8}|F_{p'_{3}}(z)|^{2}\}\label{eq4.43}
\eea
with $(D_{\alpha,\alpha,p'_{0}})^{2}\sim\epsilon$, $(D_{\alpha,\alpha,p'_{3}})^{2}\sim\frac{1}{\epsilon}$, we could suggest that if 
in series for
\beq
F_{p'_{0}}(z)=1+k_{1}z+k_{2}z^{2}+k_{3}z^{3}+k_{4}z^{4}+...\label{eq4.44}
\eeq
the coefficients $k_{1},k_{2},k_{3}$ are finite, but the coefficient $k_{4}$ is singular
\beq
k_{4}\sim\frac{1}{\epsilon}\label{eq4.45}
\eeq
due to the singular $\beta_{p'_{0}}$ coefficients of the 4th order, and, 
as a consequence,
\beq
|F_{p'_{0}}(z)|^{z}\sim\frac{1}{\epsilon^{2}}|z|^{8}\label{eq4.46}
\eeq
in this case the two terms in (\ref{eq4.43}) could compensate one another, having the appropriate coefficients.
This type of cancellation produces itself in the case of minimal models, 
in the case of resonances between the dimensions of the operators, 
inside and outside of the Kac table, correcting in this way, appropriately, the fusion rules. One particular case of the compensation of this type, of an operator outside the Kac table, by the descendent of the operator inside the table, but having the vanishing operator algebra coefficient in front, one particular case of this type of delicate decoupling is commented on in the lectures [11], Section 9.2.

We observe that the coefficients $(D_{\alpha,\alpha,p'_{0}})^{2}$ and 
$(D_{\alpha,\alpha,p'_{3}})^{2}$ need not to be positive definite. In case of minimal models, they are not positive definite, for "ghosts", 
the operators outside the Kac table.

Above is presented our prediction for the compensation between the two channels in (\ref{eq4.43}), in this way giving the same function $<\sigma\sigma\sigma\sigma>$, for the representations $\alpha_{\sigma}=\alpha_{\frac{3}{2},\frac{1}{2}}$ and $\alpha_{\sigma}=\alpha_{\frac{3}{2},\frac{3}{2}}$ of the spin operator. -- The function which is defined by the limit $\epsilon<\sigma\sigma\sigma\sigma>$, $\epsilon\rightarrow 0$.

For this prediction to be valid the necessary conditions are that the coefficients $k_{1}$, $k_{2}$, $k_{3}$ (the corresponding $\beta$ coefficients),
of the channel $p'_{0}$, should be finite, while the coefficient $k_{4}$ (the corresponding $\beta$ coefficients, coefficients of the 4th order) should diverge as $1/\epsilon$, in the regularised calculations.

These necessary conditions will be verified, below.

But we should not expect the exact compensation though, because 
the $\epsilon$ - regularisation which we shall use is not exact, as a continuation of the Potts model outside the percolation point. The leading powers of $\epsilon$ are expected to  be correct, 
but the coefficients at these powers, in particular their relative values 
of one with respect to another, 
should not  be expected to be correct, compared 
to the appropriate exact continuation of the 4-point function 
outside the percolation point, the continuation which is not known.

-- See also the corresponding remarks in the Section 3, in the text following the equations (\ref{eq3.40}), (\ref{eq3.41}). 

To summarise, we expect that, if the necessary conditions are verified, the compensation of the two channels in (\ref{eq4.43}) is in fact exact, in the proper theory.

We turn now to the regularised calculations. The modifications, to our previous calculations are limited. The finite factors, in the expressions for $(D_{\alpha,\alpha,p'_{0}})^{2}$ and $(D_{\alpha,\alpha,p'_{3}})^{2}$, will keep their values, as we are interested only in the leading order behaviour, in $\epsilon$. We have to correct, to regularise, the values of the divergent or vanishing factors only. But we will have to calculate also the $\beta$ coefficients,
of the $p'_{0}$ channel.

\underline{Channel $p'_{0} = \alpha_{1} + \alpha_{2} = 2\alpha + \epsilon$.}

We have to recalculate the factor $\Upsilon(-4\alpha_{0})$ in (\ref{eq4.16}),
originated from \\
$\Upsilon(\alpha_{1} + \alpha_{2} + p'_{0} - 2\alpha_{0})
= \Upsilon(2p'_{0} - 2\alpha_{0})$,
with $\alpha_{+}$, $\alpha_{-}$ $\epsilon$-shifted as in (\ref{eq3.36}), (\ref{eq3.37}) and $\alpha_{1}$, $\alpha_{2}$ shifted as in (\ref{eq3.40}).
In these formulas, now, $\alpha \equiv \alpha_{\sigma} 
= \alpha_{\frac{3}{2},\frac{3}{2}}$, instead of $\alpha 
= \alpha_{\frac{3}{2},\frac{1}{2}}$, in the Section 3.
We obtain ($h=\sqrt{\frac{3}{2}}$):
\bea
2\alpha_{0}
=\alpha_{-}+\alpha_{+}
\simeq-\frac{1}{h}+\frac{2}{3}\epsilon+h+\epsilon
= \frac{1}{h}(-1 + h^{2})+\frac{5}{3}\epsilon 
= \frac{1}{2h}+\frac{5}{3}\epsilon, \nn\\
-4\alpha_{0}=-\frac{1}{h}-\frac{10}{3}\epsilon\label{eq4.47}
\eea
On the other side, as $\alpha_{+}$, $\alpha_{-}$ are $\epsilon$-shifted, the modulus $h$ of the function $\Upsilon(x)$ is also shifted, $\Upsilon(x)=\Upsilon(x,\tilde{h})$. As a consequence
\beq
\Upsilon(-\frac{1}{\tilde{h}})=0\label{eq4.48}
\eeq
instead of $\Upsilon(-\frac{1}{h})=0$. As
$-\frac{1}{\tilde{h}}\simeq-\frac{1}{h}+\frac{2}{3}\epsilon$, eq.(\ref{eq3.37}), we get, in place of the factor $\Upsilon(-4\alpha_{0})$ in (\ref{eq4.16}):
\bea
\Upsilon(\alpha_{1} + \alpha_{2} + p'_{0} - 2\alpha_{0})
= \Upsilon(4\alpha + 2\epsilon - 2\alpha_{0})
=\Upsilon(- 4\alpha_{0} + 2\epsilon) \nn\\
\simeq \Upsilon(-\frac{1}{h}-\frac{10}{3}\epsilon + 2\epsilon)
= \Upsilon( - \frac{1}{h} + \frac{2}{3}\epsilon - 4\epsilon + 2\epsilon)
\simeq \Upsilon(-\frac{1}{\tilde{h}} - 2\epsilon)
\simeq -2\epsilon
\label{eq4.49}
\eea
-- for $\Upsilon$ with the modulus $\tilde{h}$.
We have used the property (B.8) of [9].

Putting the value (\ref{eq4.49}) in place of  $\Upsilon(-4\alpha_{0})$ 
in the eq.(\ref{eq4.16}), we obtain:
\beq
(D_{\alpha,\alpha,p'_{0}})^{2}=-2\epsilon\frac{h}{\gamma^{2}(\frac{3}{4})}\label{eq4.50}
\eeq

\underline{Channel $p'_{3}$.}

For this channel we have calculated, in the above, the coefficient $C_{\alpha,\alpha,p'_{3}} C_{p'^{+}_{3},\alpha,\alpha}$ in the expansion (\ref{eq4.6}), which is simpler, instead of $(D_{\alpha,\alpha,p'_{3}})^{2}$.

In the expression for $C_{\alpha,\alpha,p'_{3}}$, eq(\ref{eq4.34}), there two factors to regularise, $\gamma(3)\sim 1/\Gamma(-2)$, and $\gamma(-2)\sim\Gamma(-2)$, which originate from the factors $\gamma(2\rho)$ and $\gamma(1+\gamma+\rho)$, in (\ref{eq4.33}).

\underline{$\gamma(2\rho)$.}

\beq
\rho=\alpha_{+}^{2}=(\tilde{h})^{2}=(h+\epsilon)^{2}\simeq h^{2}+2h\epsilon=\frac{3}{2}+2h\epsilon\label{eq4.51}
\eeq
\bea
\gamma(2\rho)=\gamma(3+4h\epsilon)\simeq\frac{\Gamma(3)}{\Gamma(-2-4h\epsilon)}\nn\\
\simeq\frac{4}{\Gamma(-4h\epsilon)}\simeq-16h\epsilon\label{eq4.52}
\eea

\underline{$\gamma(1+\gamma+\rho)$.}

$p'_{3}=\alpha_{1}+\alpha_{2}+2\alpha_{-}=2\alpha+\epsilon+2\alpha_{-}$, according to the difinition of $p'_{3}$, eq.(\ref{eq4.8}), 
and the $\epsilon$ shifts of $\alpha_{1}$, $\alpha_{2}$ in (\ref{eq3.40}).
\beq
\gamma
=2\alpha_{+}p'_{3}=2\alpha_{+}(\frac{3}{2}\alpha_{-}-\frac{1}{2}\alpha_{+}+\epsilon)=-3-\rho+2h\epsilon\label{eq4.53}
\eeq
\bea
\gamma(1+\gamma+\rho)=\gamma(-2+2h\epsilon)
\simeq \frac{\Gamma(-2+2h\epsilon)}{\Gamma(3)}\nn\\
\simeq\frac{1}{4}\Gamma(2h\epsilon)
\simeq\frac{1}{8h\epsilon}\label{eq4.54}
\eea
By putting the values (\ref{eq4.52}), (\ref{eq4.54}) for $\gamma(2\rho)$, $\gamma(1+\gamma+\rho)$ into (\ref{eq4.33}), we obtain
\beq
C_{\alpha,\alpha,p'_{3}}\simeq\frac{1}{Z}\gamma(\frac{3}{2})\gamma^{2}(\frac{3}{4})\gamma^{2}(\frac{9}{4})\gamma(-\frac{7}{2})\cdot(-2)\label{eq4.55}
\eeq
--instead of (\ref{eq4.34}).

In $C_{p'^{+}_{3},\alpha,\alpha}$, eq.(\ref{eq4.38}), the factor $\gamma(-2)$ has to be regularised, originated from the factor $\gamma(1+\gamma'+\rho')$.

We find, with $p'^{+}_{3}=2\alpha_{0}-p'_{3}=-\frac{1}{2}\alpha_{-}+\frac{3}{2}\alpha_{+}-\epsilon$,
\beq
\gamma'=2\alpha_{-}p'^{+}_{3}=2\alpha_{-}(-\frac{1}{2}\alpha_{-}+\frac{3}{2}\alpha_{+}-\epsilon) 
= - \rho'-3+2\frac{1}{\tilde{h}}\epsilon
\simeq - \rho'-3+\frac{2\epsilon}{h}\label{eq4.56}
\eeq
\beq
\gamma(1+\gamma'+\rho')=\gamma(-2+\frac{2\epsilon}{h})
\simeq\frac{\Gamma(-2+\frac{2\epsilon}{h})}{\Gamma(3)}
\simeq\frac{1}{4}\Gamma(\frac{2\epsilon}{h})\simeq\frac{h}{8\epsilon}\label{eq4.57}
\eeq
Putting (\ref{eq4.57}) into (\ref{eq4.38}), we obtain:
\beq
C_{p'^{+}_{3},\alpha,\alpha}=\frac{1}{Z}\gamma(\frac{2}{3})\gamma(\frac{4}{3})\gamma^{2}(\frac{7}{6})\gamma^{2}(\frac{11}{6})\gamma(-\frac{8}{3})\times\frac{h}{8\epsilon}\label{eq4.58}
\eeq
-- instead of(\ref{eq4.38}).

With (\ref{eq4.55}) for $C_{\alpha,\alpha,p'_{3}}$ and (\ref{eq4.58}) for $C_{p'^{+}_{3},\alpha,\alpha}$ we obtain the $p'_{3}$ channel coefficient, in the expansion (\ref{eq4.6}):
\beq
C_{\alpha,\alpha,p'_{3}}C_{p'^{+}_{3},\alpha,\alpha} \propto\frac{1}{\epsilon}\label{eq4.59}
\eeq
We remind that to get the corresponding coefficient $(D_{\alpha,\alpha,p'_{3}})^{2}$, the product of $C_{\alpha,\alpha,p'_{3}}$ $C_{p'^{+}_{3},\alpha,\alpha}$ has to be multiplied by the factor $Z/(N_{\alpha})^{4}$, according to the formula (\ref{eq3.15}). The value of this factor, for $\alpha=\alpha_{\frac{3}{2},\frac{3}{2}}$, is given (\ref{eq4.12}) and in the Appendix C. 
It is finite, so that, as in (\ref{eq4.59}), 
\beq
(D_{\alpha,\alpha,p'_{3}})^{2} \propto \frac{1}{\epsilon}
\label{eq4.60} 
\eeq

Finally, to verify the necessary conditions for the cancelation of the channels $p'_{0}$ and $p'_{3}$, based on the analysis above of the expression (\ref{eq4.43}), we have to verify that, in the expansion (\ref{eq4.44}) of the conformal bloc function $F_{p'_{0}}(z)$, the first three coefficients, $k_{1}$, $k_{2}$, $k_{3}$, are finite, while $k_{4}$ diverges as $1/\epsilon$, eq.(\ref{eq4.45}).

The results of the calculation of the corresponding $\beta$ coefficients,
for the channel $p'_{0}$, are given in the Appendix B.  
Putting these values, and the values of  
$\Delta_{\alpha}$  and $\Delta_{p}$ in (\ref{eqB.25}),
into the expressions of the coefficients
$k_{1}$, $k_{2}$, $k_{3}$, $k_{4}$ in (\ref{eq3.21}),  
one obtains the following values for these coefficients, 
in the expansion (\ref{eq4.44}) :
\beq
k_{1} \simeq \frac{1}{16}, \quad
k_{2} \simeq \frac{145}{4608}, \quad
k_{3} \simeq \frac{1547}{73728}, \quad
k_{4} \simeq  - \frac{25}{9289728} \cdot \frac{1}{\varepsilon}
\eeq
We could conclude that the necessary conditions for the cancellation of the channels $p'_{0}$ and $p'_{3}$ are in fact verified.

\vskip1.5cm

{\bf Acknowledgments.}

I grateful to Marco Picco for numerous 
stimulating discussions.

\vskip1.5cm

\appendix

\section{Integral representation of the conformal block function $F_{p_{0}}(z)$.}

The conformal block function $F_{p_{0}}(z)$, in the expression (\ref{eq3.45}) for the correlation function $<\sigma(\infty)\sigma(1)\sigma(z,\bar{z})\sigma(0)>\propto G(z,\bar{z})$,  could be given by its expansion in powers of $z$, as in (\ref{eq3.22}), (\ref{eq3.47B}), by the expansion which is dictated uniquely by the conformal invariance. Alternatively, it could be given by a particular $1D$ integral, which follow from the Coulomb Gas $2D$ integral in (\ref{eq3.4}). By the methods of [8,10], by factorisation, one finds, for the conformal block function of the $p_{0}$ channel, the following $1D$ integral:
\bea
F_{p_{o}}(z)\propto (z)^{2\alpha^{2}}(1-z)^{2\alpha^{2}}\int_{1}^{\infty}du_{1}\int_{1}^{u_{1}}du_{2}\,(u_{1})^{a}(u_{1}-1)^{b}(u_{2}-1)^{b}\nn\\
\times(u_{1}-z)^{c}(u_{2}-z)^{c}\times(u_{1}-u_{2})^{g}\label{eqA.1}
\eea
Here
\bea
a=b=c=2\alpha_{-} \alpha,\quad g=2\alpha^{2}_{-}=2\rho',\nn\\
\alpha=\alpha_{\sigma}=\alpha_{\frac{3}{2},\frac{1}{2}}=-\frac{\alpha_{-}}{4}+\frac{\alpha_{+}}{4}\label{eqA.2}
\eea

The expression under the integral (including the two factors in front) corresponds to the holomorphic 
factor of the direct average of the product of vertex operators in (\ref{eq3.4}). The factor $(z)^{2\alpha^{2}}$ in front corresponds to the term
\beq
|z|^{-4\Delta_{\alpha}+2\Delta_{p_{0}}}\label{eqA.3}
\eeq
in (\ref{eq3.45}), to its holomorphic factor. In fact $(p_{0}=2\alpha)$:
\bea
-2\Delta_{\alpha}+\Delta_{p_{0}}=-2(\alpha^{2}-2\alpha\alpha_{0})+p^{2}_{0}-2p_{0}\alpha_{0}\nn\\
=-2\alpha^{2}+4\alpha\alpha_{0}+4\alpha^{2}-4\alpha\alpha_{0}=2\alpha^{2}\label{eqA.4}
\eea
If we rule ont this factor from the expression in (\ref{eqA.1}), as this 
is the case in (\ref{eq3.45}), to have finally $F_{p_{0}}(z)$ 
normalised on 1 ($F_{p_{0}}(z)\rightarrow 1$, $z\rightarrow 0$), we get the expression (not yet normalised):
\bea
F_{p_{0}}(z)\propto(1-z)^{2\alpha^{2}} \int^{\infty}_{1}du_{1}\int_{1}^{u_{1}}du_{2}(u_{1})^{a}(u_{2})^{a}(u_{1}-1)^{b}(u_{2}-1)^{b}\nn\\
\times(u_{1}-z)^{c}(u_{2}-z)^{c}\times(u_{1}-u_{2})^{2\rho'}\label{eqA.5}
\eea
To normalise, we calculate the value of the integral above for $z=0$.
\beq
\int^{\infty}_{1}du_{1}\int^{u_{1}}_{1}du_{2}(u_{1})^{a+c}(u_{2})^{a+c}(u_{1}-1)^{b}(u_{2}-1)^{b}\times(u_{1}-u_{2})^{2\rho'}\label{eqA.6}
\eeq
To put it in the standard form of the Selberg integral [12,10], we change (invert) the variables $u_{1}$, $u_{2}$ in the above, as
\beq
u_{1}\rightarrow\frac{1}{u_{1}},\quad u_{2}\rightarrow\frac{1}{u_{2}}\label{eqA.7}
\eeq
The integral in (\ref{eqA.6}) takes the form:
\bea
\int^{1}_{0}du_{2}\int_{0}^{u_{2}}du_{1}(u_{1})^{-2-a-b-c-2\rho'}(u_{2}^{-2-a-b-c-2\rho'}\nn\\
\times(1-u_{1})^{b}(1-u_{2})^{b}(u_{2}-u_{1})^{2\rho'}\label{eqA.8}
\eea
Next we shall use the value for the Selberg integral [12], rederived by a different methods in [10].
For the cas of the double integral in (\ref{eqA.8}), the general formula 
(A.36) in [10] takes the form:
\bea
\int^{1}_{0}du_{2}\int_{0}^{u_{2}}du_{1}(u_{1})^{\alpha'}(u_{2})^{\alpha'}(1-u_{1})^{\beta'}(1-u_{2})^{\beta'}(u_{2}-u_{1})^{2\rho'}\nn\\
=\frac{\Gamma(2\rho')}{\Gamma(\rho')}\times\frac{\Gamma(1+\alpha')\Gamma(1+\alpha'+\rho')\times\Gamma(1+\beta')\Gamma(1+\beta'+\rho')}
{\Gamma(2+\alpha'+\beta'+\rho')\Gamma(2+\alpha'+\beta'+2\rho')}\label{eqA.9}
\eea
$\alpha'$, $\beta'$, $\rho'$ are the parameters of the Selberg integral. For their values in (\ref{eqA.8})
\beq
\alpha'=-2-a-b-c-2\rho',\quad \beta'=b\label{eqA.10}
\eeq
we obtain:
\beq
\frac{\Gamma(2\rho')}{\Gamma(\rho')}\times\frac{\Gamma(-1-a-b-c-2\rho')\Gamma(-1-a-b-c-\rho')\Gamma(1+b)\Gamma(1+b+\rho')}
{\Gamma(-a-c-\rho')\Gamma(-a-c)}\label{eqA.11}
\eeq
The numerical values of the parameters in (\ref{eqA.2}):
\beq
\rho'=\frac{2}{3},\quad a=b=c=2\alpha_{-}(-\frac{\alpha{-}}{4}+\frac{\alpha_{+}}{4})=-\frac{1}{2}(1+\rho')=-\frac{5}{6}\label{eqA.12}
\eeq
Finally, for the normalisation integral in (\ref{eqA.11}), we obtain:
\beq
\frac{\Gamma(\frac{4}{3})}{\Gamma(\frac{2}{3})}\cdot\frac{\Gamma^{2}(\frac{1}{6})\Gamma^{2}(\frac{5}{6})}
{\Gamma(1)\Gamma(\frac{5}{3})}\label{eqA.13}
\eeq
We denote this value, of the normalisation integral,  as $N_{int}$:
\beq
N_{int}=\frac{\Gamma(\frac{4}{3})}{\Gamma(\frac{2}{3})} \cdot\frac{\Gamma^{2}(\frac{1}{6})\Gamma^{2}(\frac{5}{6})}{\Gamma(\frac{5}{3})}=\frac{3 \Gamma(\frac{4}{3}) \Gamma^{2}(\frac{1}{6})\Gamma^{2}(\frac{5}{6})}{2 \Gamma^{2}(\frac{2}{3})}\label{eqA.14}
\eeq

Returning to the eq.(\ref{eqA.5}), normalising the integral and substituting the numerical values of the parameters in (\ref{eqA.12}), we get the following expression for the function $F_{p_{0}}(z)$:
\bea
F_{p_{0}}(z)=\frac{1}{N_{int}}(1-z)^{2\alpha^{2}}\nn\\
\times\int_{1}^{\infty}du_{1}\int_{1}^{u_{1}}du_{2}[u_{1}u_{2}(u_{1}-1)(u_{2}-1)(u_{1}-z)(u_{2}-z)]^{-5/6}\times(u_{1}-u_{2})^{4/3}\label{eqA.15}
\eea
The value of the normalisation constant $N_{int}$ is given above, 
in (\ref{eqA.14}), and the value of the power of the factor $(1-z)$, in front, is given by:
\beq
2\alpha^{2}=-2\Delta_{\alpha}+\Delta_{p_{0}}=-\frac{5}{48}+\frac{5}{8}=\frac{25}{48}\label{eqfA.16}
\eeq

The integral above defines $F_{p_{0}}(z)$ for all values of $z$. This integral is convergent, at all its limits, though it converges slowly at some limits. For instance, for $u_{2}\rightarrow 1$, from above, while $u_{1}$ is far away, the integral over $u_{2}$, close to 1, behaves as
\beq
\int_{1}du_{2}(u_{2}-1)^{-5/6}\label{eqA.17}
\eeq

\section{Matrix  elements and  $\beta$ coefficients.}

{\bf Matrix elements.}

Values of the matrix elements appearing in the developpement (\ref{eq3.20}):

\underline{order 0},
\beq
<V_{\alpha}(\infty)V_{\alpha}(1)V_{p}(0)>=C_{\alpha,\alpha,p}\label{eqB.1}
\eeq

\underline{order 1},
\beq
<V_{\alpha}(\infty)V_{\alpha}(1)L_{-1}V_{p}(0)>=C_{\alpha,\alpha,p}\Delta_{p}\label{eqB. 2}
\eeq

\underline{order 2},
\beq
<V_{\alpha}(\infty)V_{\alpha}(1)L^{2}_{-1}V_{p}(0)>=C_{\alpha,\alpha,p}\Delta_{p}(\Delta_{p}+1)\label{eqB.3}
\eeq
\beq
<V_{\alpha}(\infty)V_{\alpha}(1)L_{-2}V_{p}(0)>=C_{\alpha,\alpha,p}(\Delta_{\alpha}+\Delta_{p})\label{eqB.4}
\eeq

\underline{order 3},
\beq
<V_{\alpha}(\infty)V_{\alpha}(1)L^{3}_{-1}V_{p}(0)>=C_{\alpha,\alpha,p}\Delta_{p}(\Delta_{p}+1)(\Delta_{p}+2)\label{eqB.5}
\eeq
\beq
<V_{\alpha}(\infty)V_{\alpha}(1)L_{-1}L_{-2}V_{p}(0)>=C_{\alpha,\alpha,p}(\Delta_{\alpha}+\Delta_{p})(\Delta_{p}+2)\label{eqB.6}
\eeq
\beq
<V_{\alpha}(\infty)V_{\alpha}(1)L_{-3}V_{p}(0)>=C_{\alpha,\alpha,p}(2\Delta_{\alpha}+\Delta_{p})\label{eqB.7}
\eeq

\underline{order 4},
\beq
<V_{\alpha}(\infty)V_{\alpha}(1)L^{4}_{-1}V_{p}(0)>=C_{\alpha,\alpha,p}\Delta_{p}(\Delta_{p}+1)(\Delta_{p}+2)(\Delta_{p}+3)\label{eqB.8}
\eeq
\beq
<V_{\alpha}(\infty)V_{\alpha}(1)L^{2}_{-1}L_{-2}V_{p}(0)>=C_{\alpha,\alpha,p}(\Delta_{\alpha}+\Delta_{p})(\Delta_{p}+2)(\Delta_{p}+3)  \label{eqB.9}
\eeq
\beq
<V_{\alpha}(\infty)V_{\alpha}(1)L_{-1}L_{-3}V_{p}(0)>=C_{\alpha,\alpha,p}(2\Delta_{\alpha}+\Delta_{p})(\Delta_{p}+3)\label{eqB.10}
\eeq
\beq
<V_{\alpha}(\infty)V_{\alpha}(1)L^{2}_{-2}V_{p}(0)>=C_{\alpha,\alpha,p}(\Delta_{\alpha}+\Delta_{p})(\Delta_{\alpha}+\Delta_{p}+2)\label{eqB.11}
\eeq
\beq
<V_{\alpha}(\infty)V_{\alpha}(1)L_{-4}V_{p}(0)>=C_{\alpha,\alpha,p}(3\Delta_{\alpha}+\Delta_{p})\label{eqB.12}
\eeq

\vskip0.5cm

{\bf Coefficients $\beta$.}

Equations defining the coefficients $\beta$ up to order 4 (given in a slightly more general context, of developping the product of two primery fields $\Phi_{2}(z)\Phi_{1}(0)$ towards the channel of the operator $\Phi_{p}(0)$, instead of $V_{\alpha}(z)V_{\alpha}(0)$ towards $V_{p}(0)$):

1.
\beq
\beta^{(-1)}_{p}=\frac{\Delta_{2}-\Delta_{1}+\Delta_{p}}{2\Delta_{p}}\label{eqB.13}
\eeq

2.
\beq
(\Delta_{2}-\Delta_{1}+\Delta_{p}+1)\beta_{p}^{(-1)}=2(2\Delta_{p}+1)\beta_{p}^{(-1,-1)}+3\beta^{(-2)}_{p}\label{eqB.14}
\eeq  
  
3.
\beq
(\Delta_{2}-\Delta_{1}+\Delta_{p}+2)\beta_{p}^{(-1,-1)}=6(\Delta_{p}+1)\beta_{p}^{(-1,-1,-1)}+3\beta^{(-1,-2)}\label{eqB.15}
\eeq  

4.
\beq
(\Delta_{2}-\Delta_{1}+\Delta_{p}+2)\beta_{p}^{(-2)}=2(\Delta_{p}+2)\beta_{p}^{(-1,-2)}+4\beta^{(-3)}_{p}\label{eqB.16}
\eeq

5.
\beq
(\Delta_{2}-\Delta_{1}+\Delta_{p}+3)\beta_{p}^{(-1,-1,-1)}=4(2\Delta_{p}+3)\beta_{p}^{(-1,-1,-1,-1)}+3\beta^{(-1,-1,-2)}_{p}\label{eqB.17}
\eeq  

6.
\beq
(\Delta_{2}-\Delta_{1}+\Delta_{p}+3)\beta_{p}^{(-1,-2)}=2(2\Delta_{p}+5)\beta_{p}^{(-1,-1,-2)}+4\beta^{(-1,-3)}_{p}+6\beta_{p}^{(-2,-2)}\label{eqB.18}
\eeq  

7.
\beq
(\Delta_{2}-\Delta_{1}+\Delta_{p}+3)\beta_{p}^{(-3)}=2(\Delta_{p}+3)\beta_{p}^{(-1,-3)}-3\beta^{(-2,-2)}_{p}+5\beta_{p}^{(-4)}\label{eqB.19}
\eeq  

8.
\beq
(2\Delta_{2}-\Delta_{1}+\Delta_{p})=6\Delta_{p}\beta_{p}^{(-1,-1)}+(4 \Delta_{p}+\frac{c}{2})\beta_{p}^{(-2)}\label{eqB.20}
\eeq  

9.
\bea
(2\Delta_{2}-\Delta_{1}+\Delta_{p}+1)\beta_{p}^{(-1)}=\nn\\
6(3\Delta_{p}+1)\beta_{p}^{(-1,-1,-1)}+(9+4\Delta_{p}+\frac{c}{2})\beta^{(-1,-2)}_{p}+5\beta_{p}^{(-3)}\label{eqB.21}
\eea  

10.
\bea
(2\Delta_{2}-\Delta_{1}+\Delta_{p}+2)\beta_{p}^{(-1,-1)}=\nn\\
12(3\Delta_{p}+2)\beta_{p}^{(-1,-1,-1,-1)}+(18+4\Delta_{p}+\frac{c}{2})\beta_{p}^{(-1,-1,-2)}+5\beta_{p}^{(-1,-3)}\label{eqB.22}
\eea  

11.
\bea
(2\Delta_{2}-\Delta_{1}+\Delta_{p}+2)\beta_{p}^{(-2)}=\nn\\
6(\Delta_{p}+2)\beta_{p}^{(-1,-1,-2)}+12\beta^{(-1,-3)}_{p}+(8\Delta_{p}+8+c)\beta_{p}^{(-2,-2)}+6\beta_{p}^{(-4)}\label{eqB.23}
\eea 

\underline{Comments to this system of equations.}

Equation 1 defines $\beta_{p}^{(-1)}$.

Equations 2, 8 define $\beta_{p}^{(-1,-1)}$, $\beta_{p}^{(-2)}$.

Equations 3, 4, 9 define $\beta_{p}^{(-1,-1,-1)}$, $\beta_{p}^{(-1,-2)}$, $\beta_{p}^{(-3)}$.

Equations 5, 6, 7, 10, 11  define $\beta_{p}^{(-1,-1,-1,-1)}$, $\beta_{p}^{(-1,-1,-2)}$, $\beta_{p}^{(-1,-3)}$, $\beta_{p}^{(-2,-2)}$, $\beta_{p}^{(-4)}$.

\vskip1cm

Proceeding in this way one gets the following values 
of the coefficients $\beta$, for $\Delta_{1} = \Delta_{2} 
= \Delta_{\alpha} = 5/96$  and the channel $p_{0}$, $\Delta_{p}
= \Delta_{p_{0}} = 5/8$, Section 3:
\bea
\beta_{p_{0}}^{(-1)} = \frac{1}{2}, \quad \beta_{p_{0}}^{(-1,-1)} = \frac{13}{72}, \quad
\beta_{p_{0}}^{(-2)} = 0,  \nn\\
\beta_{p_{0}}^{(-1,-1,-1)} = \frac{7}{144}, \quad \beta_{p_{0}}^{(-1,-2)} = 0, 
\quad  \beta_{p_{0}}^{(-3)} = 0,  \nn\\
\beta_{p_{0}}^{(-1,-1,-1,-1)} = \frac{433}{48384}, 
\quad \beta_{p_{0}}^{(-1,-1,-2)} = \frac{1165}{145152}, \nn\\
\beta_{p_{0}}^{(-1,-3)} = -\frac{2825}{145152}, \quad 
\beta_{p_{0}}^{(-2,-2)} =  -\frac{725}{193536}, \quad
\beta_{p_{0}}^{(-4)} = \frac{1885}{72576}
\eea

\vskip1cm

For the channel $p'_{0}$, Section 4, with the regularised values
of $\alpha_{+}$, $\alpha_{-}$ in (\ref{eq3.36}), (\ref{eq3.37}), 
of $\alpha_{1}$, $\alpha_{2}$ in (\ref{eq3.40}) 
(with $\alpha = \alpha_{\frac{3}{2},\frac{3}{2}}$, for the Section 4)
and  of $p'_{0} = \alpha_{1} + \alpha_{2} = 2\alpha + \epsilon$,  
one finds:
\beq
c \simeq -10 \varepsilon, \quad 
\Delta_{\alpha_{1}} = \Delta_{\alpha_{2}} 
\simeq \frac{5}{96} + \frac{7}{48} \varepsilon, \quad
\Delta_{p'_{0}} \simeq \frac{1}{8} + \frac{1}{4} \varepsilon
\label{eqB.25}
\eeq
In the expressions above $\varepsilon = \epsilon/h$.

Next, by calculating the $\beta$ coefficients, for the channel $p'_{0}$,
with the equations (\ref{eqB.13}) -- (\ref{eqB.23}) above, 
and with $c$, $\Delta_{1}$, $\Delta_{2}$, $\Delta_{p}$ in (\ref{eqB.25}),
one gets the following values of the $\beta$ coefficients:
\bea
\beta^{(-1)} = \frac{1}{2}, \quad 
\beta^{(-1,-1)} = \frac{1}{4} - \frac{\varepsilon}{8}, \quad
\beta^{(-2)} = - \frac{1}{48} + \frac{\varepsilon}{16}, \nn\\
\beta^{(-1,-1,-1)} = \frac{1}{12} - \frac{\varepsilon}{16}, \quad
\beta^{(-1,-2)} = - \frac{1}{96} + \frac{\varepsilon}{32}, \quad
\beta^{(-3)} = - \frac{323}{480} \varepsilon^{2}
- \frac{3391}{450}\varepsilon^{3} \nn\\
\beta^{(-1,-1,-1,-1)} = \frac{1}{3584}\cdot \frac{1}{\varepsilon}, \quad
\beta^{(-1,-1,-2)} = - \frac{13}{10752}\cdot \frac{1}{\varepsilon}, \quad
\beta^{(-1,-3)} = \frac{31}{10752}\cdot \frac{1}{\varepsilon}, \nn\\
\beta^{(-2,-2)} = \frac{25}{129024}\cdot \frac{1}{\varepsilon}, \quad
\beta^{(-4)} = - \frac{25}{7168}\cdot \frac{1}{\varepsilon}, \quad
\eea

\section{Partition function $Z$, normalisation constants $(N_{\alpha})^{2}$,
for $\alpha = \alpha_{\frac{3}{2},\frac{1}{2}}$ and  
$\alpha = \alpha_{\frac{3}{2},\frac{3}{2}}$, 
and the coefficients $(N_{\alpha})^{4}/Z$.}

We shall group in this Appendix various formulas, expressions, numerical values, which are used in the main text.

The \underline{Coulomb Gas partition function} $Z$ is given by:
\beq
Z=\Upsilon(-2\alpha_{0})\rho^{\rho-\rho'}\label{eqC.1}
\eeq
--eq.(4.11), [9]. In turn, $\Upsilon(-2\alpha_{0})$ has the value, eq.(B.17) of [9]:
\beq
\Upsilon(-2\alpha_{0})=-\frac{\rho}{(\rho-1)^{2}}\gamma(\rho)\gamma(\rho')\rho^{-\rho+\rho'}\label{eqC.2}
\eeq
so that
\beq
Z=-\frac{\rho}{(\rho-1)^{2}}\gamma(\rho)\gamma(\rho')\label{eqC.3}
\eeq
In particular, for $\rho=h^{2}=\frac{3}{2}$, $\rho'=h^{-2}=\frac{2}{3}$, we get
\beq
\Upsilon(-2\alpha_{0})=
-6 \gamma(\frac{3}{2}) \gamma(\frac{2}{3})\rho^{-\rho+\rho'}\label{eqC.4}
\eeq
\beq
Z=-6\gamma(\frac{3}{2})\gamma(\frac{2}{3})\label{eqC.5}
\eeq

\vskip0.5cm

The \underline{normalisation squared} $(N_{\alpha})^{2}$ of the vertex operator
\beq
V_{\alpha}(z,\bar{z})=e^{i\alpha\gamma(z,\bar{z})}\label{eqC.6}
\eeq
is given by the expression:
\bea
(N_{\alpha})^{2}=<I^{+}V_{\alpha}V_{\alpha}>=\frac{\Upsilon(2\alpha-2\alpha_{0})}{\Upsilon(2\alpha)\Upsilon(-2\alpha_{0})}\times\rho^{(n'-1)(1-\rho')}(\rho')^{(n-1)(1-\rho)}\nn\\
=\frac{\Upsilon(2\alpha-2\alpha_{0})}{\Upsilon(2\alpha)\Upsilon(-2\alpha_{0})}\rho^{(n'-1)(1-\rho')-(n-1)(1-\rho)}\label{eqC.7}
\eea
-- eq.(4.37) of [9]. Here the charge $\alpha$ is supposed to be of the form:
\beq
\alpha=\alpha_{n',n}=\frac{1-n'}{2}\alpha_{-}+\frac{1-n}{2}\alpha_{+}\label{eqC.8}
\eeq

\vskip0.8cm

For \underline{$\alpha=\alpha_{\frac{3}{2},\frac{3}{2}}$}, $n'-1=\frac{1}{2}$, 
$n-1=\frac{1}{2}$, $\alpha=-\frac{\alpha_{-}}{4}-\frac{\alpha_{+}}{4}$, $2\alpha=-\frac{\alpha_{-}}{2}-\frac{\alpha_{+}}{2}=-\alpha_{0}=-\frac{1}{4h}$, we obtain:
\bea
(N_{\alpha_{\frac{3}{2},\frac{3}{2}}})^{2}=\frac{\Upsilon(-3\alpha_{0})}{\Upsilon(-\alpha_{0})\Upsilon(-2\alpha_{0})}\rho^{\frac{1}{2}(1-\rho')-\frac{1}{2}(1-\rho)}\nn\\
=\frac{\Upsilon(-3\alpha_{0})}{\Upsilon(-\alpha_{0})}\times
\frac{1}{(-6)\gamma(\frac{3}{2})\gamma(\frac{2}{3})\rho^{-\rho+\rho'}}\rho^{\frac{1}{2}(\rho-\rho')}\nn\\
=\frac{\Upsilon(-\frac{3}{4h})}{\Upsilon(-\frac{1}{4h})}\times\frac{1}{(-6)\gamma(\frac{3}{2})\gamma(\frac{2}{3})}\rho^{5/4}\label{eqC.9}
\eea
We shall use now the value of the ratio $\Upsilon(-\frac{3}{4h})/\Upsilon(-\frac{1}{4h})$ which will be obtained slightly below:
\beq
\frac{\Upsilon(-\frac{3}{4h})}{\Upsilon(-\frac{1}{4h})}=\gamma(\frac{3}{4})\rho^{-1/4}\label{eqC.10}
\eeq
Putting (\ref{eqC.10}) into (\ref{eqC.9}) we obtain:
\beq
(N_{\alpha_{\frac{3}{2},\frac{3}{2}}})^{2}=\frac{\gamma(\frac{3}{4})}{-6\gamma(\frac{3}{2})\gamma(\frac{2}{3})}\times\rho\label{eqC.11}
\eeq
As $\rho=3/2$,
\beq
(N_{\alpha_{\frac{3}{2},\frac{3}{2}}})^{2}=\frac{\gamma(\frac{3}{4})}{-4\gamma(\frac{3}{2})\gamma(\frac{2}{3})}\label{eqC.12}
\eeq
For the coefficient $(N_{\alpha})^{4}/Z$ in eq.(\ref{eq3.15}) we find
\bea
\frac{(N_{\alpha_{\frac{3}{2},\frac{3}{2}}})^{4}}{Z}=\frac{\gamma^{2}(\frac{3}{4})}{16\gamma^{2}(\frac{3}{2})\gamma^{2}(\frac{2}{3})}\times\frac{1}{(-6)\gamma(\frac{3}{2})\gamma(\frac{2}{3})}
=-\frac{1}{96}\cdot\frac{\gamma^{2}(\frac{3}{4})}{\gamma^{3}(\frac{3}{2})\gamma^{3}(\frac{2}{3})}\label{eqC.13}
\eea

\vskip0.8cm

In case of \underline{$\alpha=\alpha_{\frac{3}{2},\frac{1}{2}}$}
$=-\frac{\alpha_{-}}{4}+\frac{\alpha_{+}}{4}$, the value of $(N_{\alpha})^{2}$ is different.

We observe  that, as is easy to check,
\beq
\alpha_{\frac{3}{2},\frac{1}{2}}=\alpha^{+}_{\frac{3}{2},\frac{3}{2}}=2\alpha_{0}-\alpha_{\frac{3}{2},\frac{3}{2}}\label{eqC.14}
\eeq
which is only valid in the case of particular values of  $\alpha_{+}$
and $\alpha_{-}$: \,
$\alpha_{+} \equiv h =\sqrt{\frac{3}{2}}$, \,\,$\alpha_{-} \equiv -1/h
=-\sqrt{\frac{2}{3}}$;
in particular,  $\alpha_{0}=\frac{1}{2}(\alpha_{-}+\alpha_{+})=\frac{1}{\sqrt{24}}=\frac{1}{4h}$. 

Then, as
\beq
N_{\alpha^{+}}=\frac{1}{Z\cdot N_{\alpha}}\label{eqC.15}
\eeq
-- eq.(4.34) of [9], we obtain:
\bea
(N_{\alpha_{\frac{3}{2},\frac{1}{2}}})^{2}=
(N_{\alpha^{+}_{\frac{3}{2},\frac{3}{2}}})^{2}=
\frac{1}{Z^{2}(N_{\alpha_{\frac{3}{2},\frac{3}{2}}})^{2}}\nn\\
=\frac{1}{36\gamma^{2}(\frac{3}{2})\gamma^{2}(\frac{2}{3})}\times\frac{(-4)\gamma(\frac{3}{2})\gamma(\frac{2}{3})}{\gamma(\frac{3}{4})},\nn\\
(N_{\alpha_{\frac{3}{2},\frac{1}{2}}})^{2}=-\frac{1}{9}\cdot\frac{1}{\gamma(\frac{3}{4})\gamma(\frac{3}{2})\gamma(\frac{2}{3})}\label{eqC.16}
\eea
For the coefficient $(N_{\alpha})^{4}/Z$ we find:
\bea
\frac{(N_{\alpha_{\frac{3}{2},\frac{1}{2}}})^{4}}{Z}=\frac{1}{81}\cdot\frac{1}{\gamma^{2}(\frac{3}{4})\gamma^{2}(\frac{3}{2})\gamma^{2}(\frac{2}{3})}\times\frac{1}{(-6)\cdot\gamma(\frac{3}{2})\gamma(\frac{2}{3})}\nn\\
\frac{(N_{\alpha_{\frac{3}{2},\frac{1}{2}}})^{4}}{Z}=-\frac{1}{486}\cdot\frac{1}{\gamma^{2}(\frac{3}{4})\gamma^{3}(\frac{3}{2})\gamma^{3}(\frac{2}{3})}\label{eqC.17}
\eea

\vskip0.5cm

We shall justify now the value in (\ref{eqC.10}) for the ratio 
$\Upsilon(-\frac{3}{4h})/ \Upsilon(-\frac{1}{4h})$, $\Upsilon(x)$ having 
the modulus $h$, $\Upsilon(x)\equiv\Upsilon(x,h)$, with for $h=\sqrt{3/2}$.

This could be derived as follows.
\beq
\Upsilon(-\frac{3}{4h})=
\Upsilon(-\frac{3}{4h^{2}}h)=
\Upsilon(-\frac{3}{4}\cdot\frac{2}{3}\cdot h)=
\Upsilon(-\frac{h}{2})=
\Upsilon(\frac{h}{2}-h)\label{eqC.18}
\eeq
Next we shall use the formula (B.5),[9]
\beq
\Upsilon(x-h)=\gamma(-h(x-h))h^{1+2h(x-h)}\times\Upsilon(x)\label{eqC.19}
\eeq
with $x=\frac{h}{2}$. We obtain:
\bea
\Upsilon(-\frac{3}{4h})=\Upsilon(\frac{h}{2}-h)=\gamma(-h(-\frac{h}{2}))h^{1+2h\cdot(-\frac{h}{2})}\times\Upsilon(\frac{h}{2})\nn\\
=\gamma(\frac{3}{4})h^{-1/2}\Upsilon(\frac{h}{2})=\gamma(\frac{3}{4})h^{-1/2}\Upsilon(2\alpha_{0}-\frac{h}{2})\nn\\
=\gamma(\frac{3}{4})h^{-1/2}\Upsilon(\frac{1}{2h}-\frac{h}{2})=\gamma(\frac{3}{4})h^{-1/2}\Upsilon(\frac{1}{2h}(1-h^{2}))\nn\\
=\gamma(\frac{3}{4})h^{-1/2}\Upsilon(-\frac{1}{4h})\label{eqC.20}
\eea
In the above, we have used the property $\Upsilon(x)=\Upsilon(2\alpha_{0}-x)$ of the $\Upsilon$ function, and the specific value of $h$, several times.

From (\ref{eqC.20}) we obtain the value of the ratio in (\ref{eqC.10}), $h=\sqrt{\rho}$.

Finally we shall justify the statements, used in the analysis in Section 4, that $\Upsilon(-\alpha_{0})=\Upsilon(-\frac{1}{4h})$ and $\Upsilon(-3\alpha_{0})=\Upsilon(-\frac{3}{4h})$ have finite values.

In fact, the integral which defines the $\log\Upsilon(x,h)$
\beq
\log\Upsilon(x,h)=\int_{0}^{\infty}\frac{dt}{t}\{(\alpha_{0}-x)^{2}e^{-t}-\frac{\sinh^{2}[\alpha_{0}-x)\frac{t}{2}]}{\sinh\frac{t}{2h}\cdot\sinh\frac{ht}{2}}\}\label{eqC.21}
\eeq
is convergent at $t\rightarrow 0$ for all valus of $x$, $\alpha_{0}$, while at $t\rightarrow\infty$, the integral, of the second term in (\ref{eqC.21}), diverges at $t\rightarrow-\frac{1}{h}$, from above, and at $t \rightarrow h$, 
from below.  But it is convergent for $x$ everywhere 
in the interval $(-\frac{1}{h},h)$. So it is convergent at $x=-1/4h$. As a consequence, $\Upsilon(-1/4h,h)$ has finite and non-zero value.

$\Upsilon(-\frac{3}{4h},h)$ is finite, and non-zero, for the same reason, $-\frac{3}{4h}$ is located inside of the interval $(-\frac{1}{h},h)$. Or, otherwise, because the ratio in (\ref{eqC.10}) is finite.

\end{document}